\documentclass[amsmath,amssymb,superscriptaddress, showpacs, showkeys,twocolumn,prb]{revtex4-1}

\usepackage{graphicx}
\usepackage{dcolumn}
\usepackage{bm}
\usepackage{amsmath}
\usepackage{amssymb,amsthm}
\usepackage{color}
\usepackage{epstopdf} 
\usepackage[sans]{dsfont}
\usepackage{accents}
\usepackage[utf8]{inputenc}
\usepackage[T1]{fontenc}
\usepackage[english]{babel}
\usepackage{hyperref}
\usepackage{mathbbol,bbm}
\usepackage{float}
\usepackage{accents}
\usepackage{mathtools} 
\usepackage{booktabs}

\newcommand{\tr}[1]{\ensuremath{\text{Tr}\left[#1\right]}}

\def\Im{\text{Im}}

\def\ket#1{\mathinner{|{#1}\rangle}}

\newcommand{\ketbra}[2]{\left|#1\rangle\langle #2\right|}

\newcommand{\matelem}[3]{\langle #1|#2|#3\rangle}

\newcommand{\gin}{\textrm{in}}
\newcommand{\gout}{\textrm{out}}
\newcommand{\grec}{{\textrm{loss}}}
\newcommand{\st}{{\textrm{st}}}
\newcommand{\Imax}{{se}}
\newcommand{\crit}[1]{\tilde{#1}}

\newcommand{\ch}[1]{#1}

\begin{document}

\title{Towards high-temperature coherence-enhanced transport in few-atomic layers heterostructures}

\author{Chahan M. Kropf} 
\email{Chahan.Kropf@pv.infn.it}
\affiliation{Istituto Nazionale di Fisica Nucleare, Sezione di Pavia, via Bassi 6, I-27100 Pavia, Italy
}
\affiliation{Department of Physics, Universit\`a Cattolica del Sacro Cuore, Brescia I-25121, Italy}
\affiliation{ILAMP (Interdisciplinary Laboratories for Advanced Materials Physics), Universit\`a Cattolica del Sacro Cuore, Brescia I-25121, Italy} 
\author{Angelo Valli}
\affiliation{Scuola  Internazionale  Superiore  di  Studi  Avanzati  (SISSA), and CNR-IOM DEMOCRITOS, Istituto Officina dei Materiali, Consiglio Nazionale delle Ricerche, Via  Bonomea  265, I-34136  Trieste,  Italy}
\author{Paolo Franceschini}
\affiliation{Department of Physics, Universit\`a Cattolica del Sacro Cuore, Brescia I-25121, Italy}
\affiliation{ILAMP (Interdisciplinary Laboratories for Advanced Materials Physics), Universit\`a Cattolica del Sacro Cuore, Brescia I-25121, Italy}
\affiliation{Department of Physics and Astronomy, KU Leuven, Celestijnenlaan 200D, 3001 Leuven, Belgium}
\author{Giuseppe Luca Celardo}
\affiliation{Benem\'erita Universidad Aut\'onoma de Puebla, Apartado Postal J-48, Instituto de F\'isica, 72570, Mexico}
\author{Massimo Capone}
\affiliation{Scuola  Internazionale  Superiore  di  Studi  Avanzati  (SISSA), and CNR-IOM DEMOCRITOS, Istituto Officina dei Materiali, Consiglio Nazionale delle Ricerche, Via  Bonomea  265, I-34136  Trieste,  Italy}
\author{Claudio Giannetti} 
\affiliation{Department of Physics, Universit\`a Cattolica del Sacro Cuore, Brescia I-25121, Italy}
\affiliation{ILAMP (Interdisciplinary Laboratories for Advanced Materials Physics), Universit\`a Cattolica del Sacro Cuore, Brescia I-25121, Italy}
\author{Fausto Borgonovi} 
\email{Fausto.Borgonovi@unicatt.it}
\affiliation{Istituto Nazionale di Fisica Nucleare, Sezione di Pavia, via Bassi 6, I-27100 Pavia, Italy
}
\affiliation{Department of Physics, Universit\`a Cattolica del Sacro Cuore, Brescia I-25121, Italy}
\affiliation{ILAMP (Interdisciplinary Laboratories for Advanced Materials Physics), Universit\`a Cattolica del Sacro Cuore, Brescia I-25121, Italy} 



\date{\today}

\begin{abstract}
\noindent
The possibility to exploit quantum coherence to strongly enhance the efficiency of charge transport in solid state devices working at ambient conditions would pave the way to disruptive technological applications. In this work, we tackle the problem of the quantum transport of photogenerated electronic excitations subject to dephasing and on-site Coulomb interactions. We show that the transport to a continuum of states representing metallic collectors can be optimized by exploiting the "superradiance" phenomena. We demonstrate that this is a coherent effect which is robust against dephasing and electron-electron interactions in a parameters range that is compatible with actual implementation in few monolayers transition-metal-oxide (TMO) heterostructures.
\end{abstract}

\maketitle


\section{Introduction}
Manipulating the quantum coherence of electronic wavefunctions on timescales compatible with every-day technology is one of the big challenges of condensed matter physics \cite{Streltsov2017}. The possibility to generate, control and collect charges without loosing the information stored in the quantum phase would pave the way to the design of novel building blocks for quantum computation or optoelectronic devices with unprecedented performances based on quantum effects \cite{Awschalom2007,Datta2005}. 

One of the most striking examples of coherence-assisted processes is related to transport of charges or excitons in open quantum chains \cite{Eisert2015}. When the electronic coherence is preserved on timescales longer than the  transfer time of the charges to external collectors, new phenomena with no classical counterpart emerge. Paradigmatic examples include the transport enhancement in the vicinity of the \textit{superradiant} transition in systems coupled to one or several decay channels \cite{Celardo2009}, or the central-symmetry enhanced exciton transfer in dipole-networks in the presence of a dominant-doublet \cite{Walschaers2013}.

The concept of coherence-assisted transport has been widely employed to address the hitherto unexplained efficiency of light-harvesting processes in molecular bio-complexes \cite{Scholes2011,Huelga2013}. The experimental claim that electronic coherence is preserved on timescales of the order of several hundreds of femtoseconds has triggered a huge effort to understand and model the interaction between the photo-injected excitons and the ambient-temperature bath, mainly constituted by the excitation of the vibrational degrees of freedom of the molecules along the transport path. 

Extending this  phenomenon to solids has been so far believed possible only at extremely low temperatures \cite{Beenakker1991} or in extremely controlled artificial solids, such as cold atoms trapped in perfect optical lattices \cite{Chien2015,Araujo2016}. This conventional wisdom is based on the assumption that the quantum nature of the electronic excitations in solids at high temperatures is washed out within few femtoseconds by the interaction with the incoherent fluctuations of the lattice, spin and charge degrees of freedom, acting as a bath.
Nonetheless, the recent advances in the synthesis of materials as thin as few atomic layers \cite{Hwang2012} opened the path to the nano-engineering of devices working in a parameter range not far from what is needed to observe and exploit coherent phenomena \cite{Gandolfi2017}. In this work we provide strong arguments to support the conclusion that signatures of coherence-enhanced transport can survive in solids and impact the efficiency of properly tailored few-monolayers transition-metal-oxide (TMO) devices.

\begin{figure}
	\centering
	\includegraphics[width = \columnwidth]{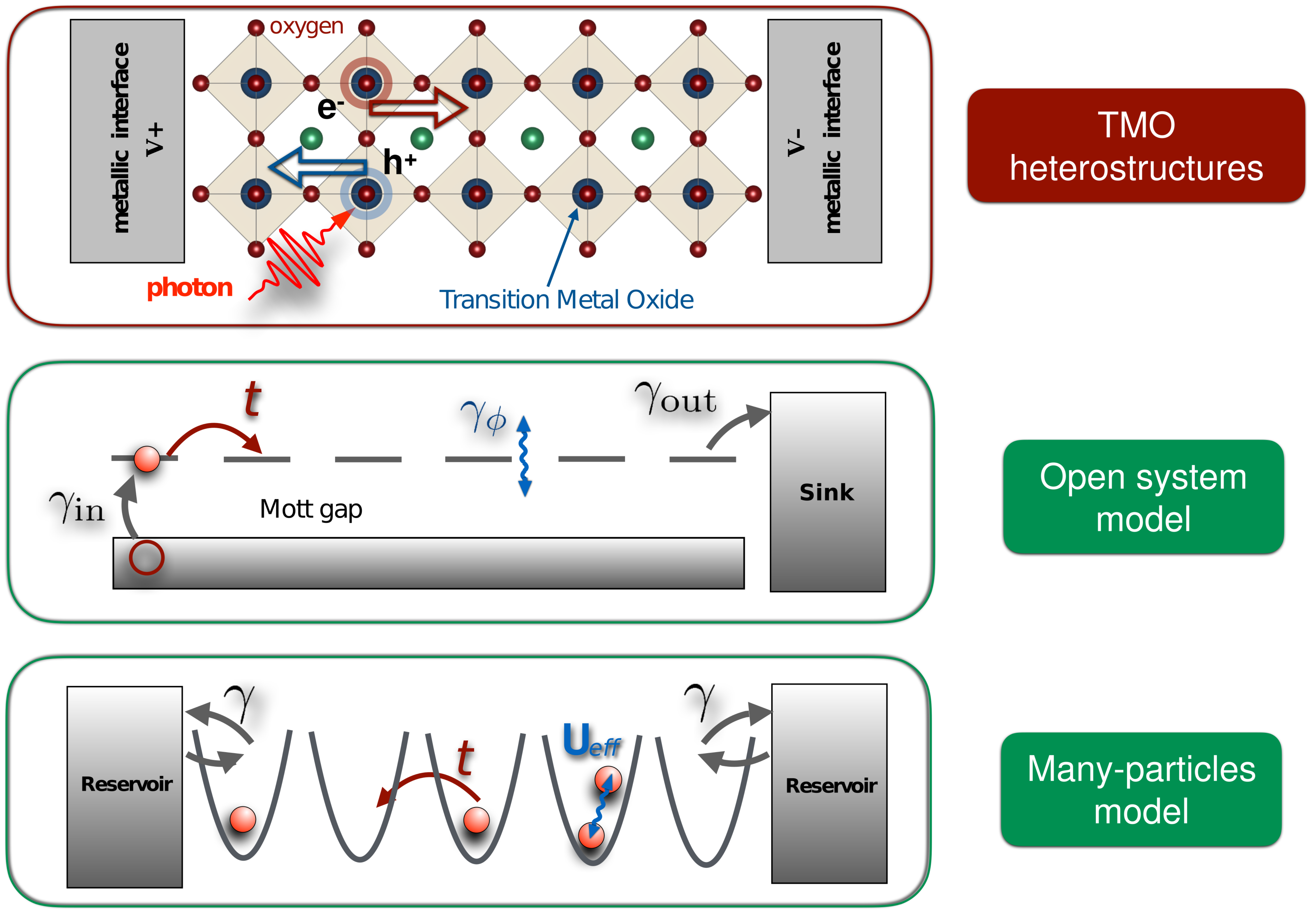}
	\caption{[Color online] Illustration of the charge generation and collection along quantum coherent pathways in a few-monolayers oxide heterostructure (top panel). The process is modelled out-of-equilibrium via an effective one-dimensional open quantum system for a single excitation (middle panel) in Section \ref{sec:master_eq} and in-equilibrium via an effective many-particle Hubbard-model (lower panel) in Section \ref{sec:DMFT}.}
\label{fig:hetero}
\end{figure}

The model example TMO that serves as motivation to the present work is qualitatively illustrated in Fig.~\ref{fig:hetero}. We consider a device constituted by $N$ atomic layers of a solid-state insulating perovskite coupled to isostructural conductive layers (collectors) at the two edges. Such a device can be realized with a very small degree of intrinsic disorder by interfacing TMOs with different degrees of electronic correlations. While the core of the device is constituted by a Mott insulator with a non-zero energy gap ($E_{\mathrm{gap}}$), the two metallic overlayers can be used to collect the photogenerated charges. The electron-hole excitation generated by the absorption of an above-gap photon ($\hbar\omega>E_{\mathrm{gap}}$) can thus hop across the TMO layers under the influence of either an applied external voltage or the intrinsic field in the case of a polar TMO. As a consequence of the strong oxygen-mediated hybridization of the orbitals connecting two adjacent atomic layers, the interlayer hopping integral ($t$), can be as large as 100-200 meV \cite{DeRaychaudhury2007}, which corresponds to a timescale of 3-6 femtoseconds ($\hbar$=658 meV fs). The timescale of the charge migration to one of the two collectors is thus confined to few femtoseconds, which is very close to the typical timescale of the electronic decoherence related to the interaction with phonons, spin and charge fluctuations. Furthermore, the large flexibility in the growth of highly controlled interfaces allows to properly tune the interface hopping integral and, consequently, engineer the escape probability in order to optimize coherent transport phenomena. At the same time, the presence of multiple structural, magnetic and orbital ordering phase transitions constitutes a fertile playground to investigate the way lattice, spin and electronic fluctuations affect the decoherence processes and suppress or favor possible quantum-transport phenomena.

In this work, we set the basic theoretical framework to treat the possible emergence of  quantum transport effects in real heterostructures at ambient conditions. We believe that, before entering into a full-fledged, realistic simulation of the non-equilbrium dynamics of photoexcited carriers in the heterostructure, it is crucial to tackle the problem by introducing simplified models which contain the basic bricks of our understanding. We start from an idealized description of the heterostructure where the interlayer transport of a single excitation is modeled by a one dimensional $N$-site chain connected to one collector and one source. Within this simple approach we obtain analytic expressions to estimate the parameters range compatible with quantum-transport efficiency enhancement, and we are able to connect the optimal transport conditions with the superradiant transition\cite{Sokolov1992,Sokolov1988,Celardo2010}, which takes place as a function of the ratio between the coupling with the leads $\gamma_{out}$ and the hopping between the sites of the chain $t$.
When $\gamma_{out} \ll t$ all the $N$ eigenstates acquire a finite lifetime of the order of $N$/$\gamma_{out}$, thus allowing coherent transport throughout the chain. In the opposite regime $\gamma_{out} \gg t$ one \textit{superradiant} edge solution appears, spatially located at the edge ($N^{th}$) site, which acquires a broadening much larger than the average one. In this regime, the lifetime (inverse broadening) of the remaining $N$-1 \textit{subradiant} solutions progressively increases thus suppressing transport at very large couplings.

The optimal regime for quantum transport where the current is maximized is reached in the vicinity of the superradiant solution separating the two regimes. The inherent quantum nature of this phenomenon can be appreciated coupling the whole chain with an external environment which leads to dephasing. As the dephasing rate becomes comparable to the inverse transfer time, the superradiance-driven current increase progressively vanishes until an incoherent hopping regime is reached. In this scenario, which applies to almost all conventional devices, the charge collection is based on diffusive or electric-field assisted charge migration at the effective drift velocity $\overrightarrow{v_\mathrm{D}}$. 

The description of a photoinjected excitation as a free particle is a useful starting point, but it completely neglects the presence of the other carriers and the electron-electron interactions, which are by no means negligible in TMOs. In the second part of our theoretical analysis we release this approximation by considering the photoexcitation as a quasiparticle in a strongly interacting environment. To treat this case we consider a simplified single-band Hubbard model in the strongly-correlated metal regime, and we show, within linear-response theory, that optimal transport occurs at the same superradiant transition even in presence of strong interaction.

The occurrence of similar coherence-driven phenomena in the two models, which describe a priori different physics, is instructive. In the first model we study non-equilibrium dynamics of a free single excitation, while in the second we compute the transmission of interacting many-particles within a Mott insulator at equilibrium. Since we obtain in both cases that optimal transport occurs at the superradiant transition, we argue that the underlying coherent quantum transport mechanism is an ubiquitous and robust mechanism observable and exploitable in actual devices at small length and time scales.
In other words, in order to see superradiant-enhancement of the charge transport efficiency we must work in a regime far from the diffusive bulk transport characterized by the drift velocity (c.f. Appendix \ref{app:ConventionalMaterials}). Nevertheless, by allowing moderate levels of dephasing, disorder, interaction and loss of charges, we are not limited to the pure ballistic transport that requires full quantum control. Finally, we will address the paradigmatic case of LaVO$_3$/SrVO$_3$ heterostructures, in which the emergence and exploitation of this coherence-enhanced transport is experimentally within reach.

The work is organized as follows. In Section \ref{sec:master_eq} we present the effective tight-binding quantum-transport model in which the coupling to the environment (lattice, spin, charge fluctuations) is described effectively within the quantum master equation formalism. We find an analytical formulation of the non-equilibrium steady state current. We show that the current is maximized when the coupling to the sink is close to the superradiant transition, whereas it is strongly suppressed when coherence is destroyed by environment-induced dephasing. We identify the dephasing threshold and we address the maximal length-scale compatible with coherence-enhanced transport. Finally, we discuss the effect of static disorder and losses on the current. In Section \ref{sec:DMFT} we benchmark our simple model against dynamical mean-field theory calculations (DMFT), which fully treat the on-site Coulomb repulsion and allows us to compute the conductance in a strongly interacting system. We show that the coherence-driven enhancement of the current flowing across the device is robust against electronic correlations. In Sec. \ref{sec:LVO} we explicitly discuss the possible realization of a coherence-enhanced device in SrVO$_3$/LaVO$_3$/SrVO$_3$ heterostructures \cite{Assmann2013}.


\section{Open system transport model}
\label{sec:master_eq}
As a first step, we model the transport of photoexcitations within a heterostructure constituted by $N$ monolayers by means of a tight-binding Hamiltonian for a chain of $N$ sites coupled with nearest-neighbour interactions (see the open system model illustration in Fig.~\ref{fig:hetero}). We will restrict our considerations to the single excitation subspace which is a good approximation when describing photo-induced current with a low rate of incident photons, such as is the case for natural sunlight \cite{Brumer2018} (see also Sec. \ref{sec:LVO}). The injection and extraction of a charge excitation, as well as the effective dephasing and decay processes, are taken into account by coupling ('opening') the system to an environment \cite{Gurvitz1996}. The excitations are injected on the left and extracted on the right of the chain as illustrated in Fig.~\ref{fig:hetero}. In this way we model effectively the directionality induced by the intrinsic electric field across TMO heterostructures which drives the positive (negative) charges towards the one negative (positive) metallic interface at the end of the chain. In principle, excess charges can be photo-generated anywhere in the heterostructure, and thus the effective chain length (i.e., the number of hopping events required to reach the metallic interface) varies from one photon absorption event to another. However, since we work in the low fluence regime the absorption events can be considered independent (single-excitation approximation). Thus, we can optimize the current for any fixed length $N$ of the chain.

Mathematically we describe the dynamics with a Lindblad master equation \cite{Lindblad1976,Gorini1976} which is a good approximation when the correlations in the environment decay much faster than the system's time-scale,
\begin{align}\label{eq:tight-binding equation}
	\dot{\hat{\rho}} = -\frac{i}{\hbar} \left[\hat{\textrm{H}}, \hat{\rho}\right] + \sum_\alpha \mathcal{L}_\alpha (\hat{\rho}).
\end{align}
This is generically the case for the system we consider (see Section \ref{sec:LVO}). In Eq.~\eqref{eq:tight-binding equation} the superoperator is $\mathcal{L}_\alpha (\hat{\rho}) = \hat{L}_\alpha \hat{\rho} \hat L_\alpha^\dagger - \frac{1}{2} \left(\hat L_\alpha^\dagger \hat L_\alpha\hat\rho + \hat\rho \hat L_\alpha^\dagger \hat L_\alpha\right)$,  where $\hat L_\alpha$ are the Lindblad operators including the corresponding rates that we introduce below, and $\hat\rho$ is the one-particle density matrix.

The Hamiltonian in Eq.~\eqref{eq:tight-binding equation} reads
\begin{align}\label{eq:hamiltonian}
	\textrm{H} = &\sum_{j = 1} ^ N \epsilon_j \ketbra{j}{j} + t \sum_{j = 1} ^ {N-1} \Big( \ketbra{j}{j+1}  +\ketbra{j+1}{j} \Big), 
\end{align}
where $\epsilon_j$ are the on-site energies and $t$ is the homogeneous hopping energy. Here $\ketbra{j}{j}$ denotes the state when one excitation is present on site $j$, and, in general, we will assume degenerate on-site energies $\epsilon_j =\epsilon = 0$. Static disorder can be eventually added by choosing  non-homogeneous on-site energies $\epsilon_j$. For example, Anderson-type disorder \cite{Anderson1958} will be introduced assuming  $\epsilon_j$ randomly distributed  in  $\left[-W/2, W/2\right]$, where $W$ is the disorder strength.
Realistic level of disorder weakly affects our system and it is thus neglected at a first stage. In Appendix \ref{app:Disorder} a detailed analysis of the effect of disorder is presented. 

The contributions arising from the coupling to the environment are taken into account by the Lindblad operators as follows: the injection of an excitation and the coupling to the exit lead consists of a pump coupled to the first site (with the rate in, $\gamma_\gin$) and a sink coupled to the last ($N$-th) site (with the rate out, $\gamma_\gout$), respectively,
\begin{align}\label{eq:LopInOut}
	&\hat L_\gin = \sqrt{\gamma_\gin}~\ketbra{1}{0} &\hat L_\gout = \sqrt{\gamma_\gout}~ \ketbra{0}{N}.
\end{align}
Here $\ketbra{0}{0}$ denotes the state when no excitation is present in the system. Dynamical noise from the coupling to the environment is modelled by uniform dephasing \cite{Haken1973} with a rate $\gamma_\phi$ on each site, which can be described by $N$ Lindblad operators 
\begin{align}\label{eq:LopDeph}
	\hat L_{\phi,j} = \sqrt{\gamma_\phi}~\ketbra{j}{j} \;\; \textrm{for} \;\; j = 1,\ldots,N.
\end{align}
The loss of excitations in the system (due to fluorescence or electron-hole recombination for instance) is taken into account by a spontaneous decay to the empty state (recombination) on each site via the $N$ identical Lindblad operators
\begin{align}\label{eq:LopRec}
		\hat L_{\grec,j} = \sqrt{\gamma_\grec} \ketbra{0}{j} \;\; \textrm{for} \;\; j = 1, \ldots, N.
\end{align}
The transport shall be characterized in terms of the non-equilibrium steady-state (NESS) probability current \cite{feist2015,Manzano2012a} leaving the system through the sink attached to the last site $N$,
\begin{align}\label{eq:current def}
	I := \gamma_\gout \matelem{N}{\hat{\rho}_\textrm{ss}}{N},
\end{align}
where $\hat{\rho}_\textrm{ss}$ is the steady-state density matrix.

In order to understand the properties of the current it is useful to consider also the transient dynamics in addition to the steady-state properties. This can be realized by setting the pumping rate $\gamma_\gin =0$. The dynamics is then obtained by solving in time the Lindblad equation \eqref{eq:tight-binding equation} for a single excitation initially located on one site, or distributed over many sites. Note that for $\gamma_\gout \neq 0$ the excitation will asymptotically leave the system.
Solving dynamically the Lindblad equation allows to find  the average transfer time to the sink, i.e. the average time the excitation needs to exit the system through the sink. This quantity is a function of the initial state, the dephasing rate, the coupling to the leads, the length of the chain, and the loss rate. \ch{The average transfer time \cite{Rebentrost2009a,Mohseni2008} can be defined via  the probability current $I(t^{\prime}) := \gamma_\gout \matelem{N}{\hat \rho (t^{\prime})}{N}$ escaping through the sink (given an initial state $\hat\rho(t_0=0)$, c.f. Appendix \ref{app:NESScurrent}), via the time integral:}
\begin{align}\label{eq:TauDef}
	\tau := \frac{1}{\eta} \int_0^\infty t^{\prime} I(t^{\prime}) dt^{\prime} = \frac{\gamma_\gout}{\eta}\int_0^\infty t^{\prime} \matelem{N}{\hat \rho (t^{\prime})}{N} dt^{\prime},
\end{align}
where we introduced the normalization through the efficiency
\begin{align}\label{eq:Efficiency}
	\eta := \gamma_\gout \int_0^\infty I(t^{\prime}) dt^{\prime}.
\end{align}
In practice, the time integrals in Eqs.~\eqref{eq:TauDef} and \eqref{eq:Efficiency} can be solved analytically in terms of the eigenvalues of the system's Liouvillian, as shown in Appendix \ref{app:AnalyticTransferEfficiency}. \ch{Note that here and throughout the rest of this manuscript we use $t^{\prime}$ as the time-variable to avoid confusion with the hopping parameter $t$.}

\subsection{Coherent transport and optimal current}
We studied the NESS current \eqref{eq:current def} numerically using the steady-state solver of the Qutip \cite{Johansson2012,Johansson2013} Python package. Unless specified otherwise, we set $N=10, \gamma_\grec =0, W=0$ (the cases $W\neq0$, $\gamma_\grec\neq0$ are discussed in Appendix \ref{app:Disorder}). In the following we show that, at low dephasing, the coupling to the sink $\gamma_\gout$ can be tuned to maximize the current.
This is consistent with results found in previous work on coherent quantum transport in the photosynthetic molecular complexes \cite{Celardo2012}. Moreover, these optimal parameters are robust against the introduction of short-range Coulomb interaction and are compatible with experimental data for chains made of TMOs atomic layers, as we shall discuss in Sections \ref{sec:DMFT} and \ref{sec:LVO}, respectively. To begin with, let us briefly explain how the different parameters of the model affect qualitatively the current. 

Our model is characterized by five parameters: the hopping  $t$, which we set as the reference energy scale, and the rates $\gamma_\gout$, $\gamma_\phi$, $\gamma_\grec$, and $\gamma_\gin$ (note that the value of $\epsilon_0$ does not affect the current). While the first three rates are fixed by the system (i.e., material) properties, the rate $\gamma_\gin$ is in general an external, tunable parameter. For instance when describing photo-induced current, $\gamma_\gin$ characterizes the rate of generated excitation from incoming photons. As we will see below, in the single excitation approximation, the current is proportional to $\gamma_\gin$, and saturates at a maximum value defined by the material parameters. 

\begin{figure}
	\centering
	\includegraphics[width = 0.95\columnwidth]{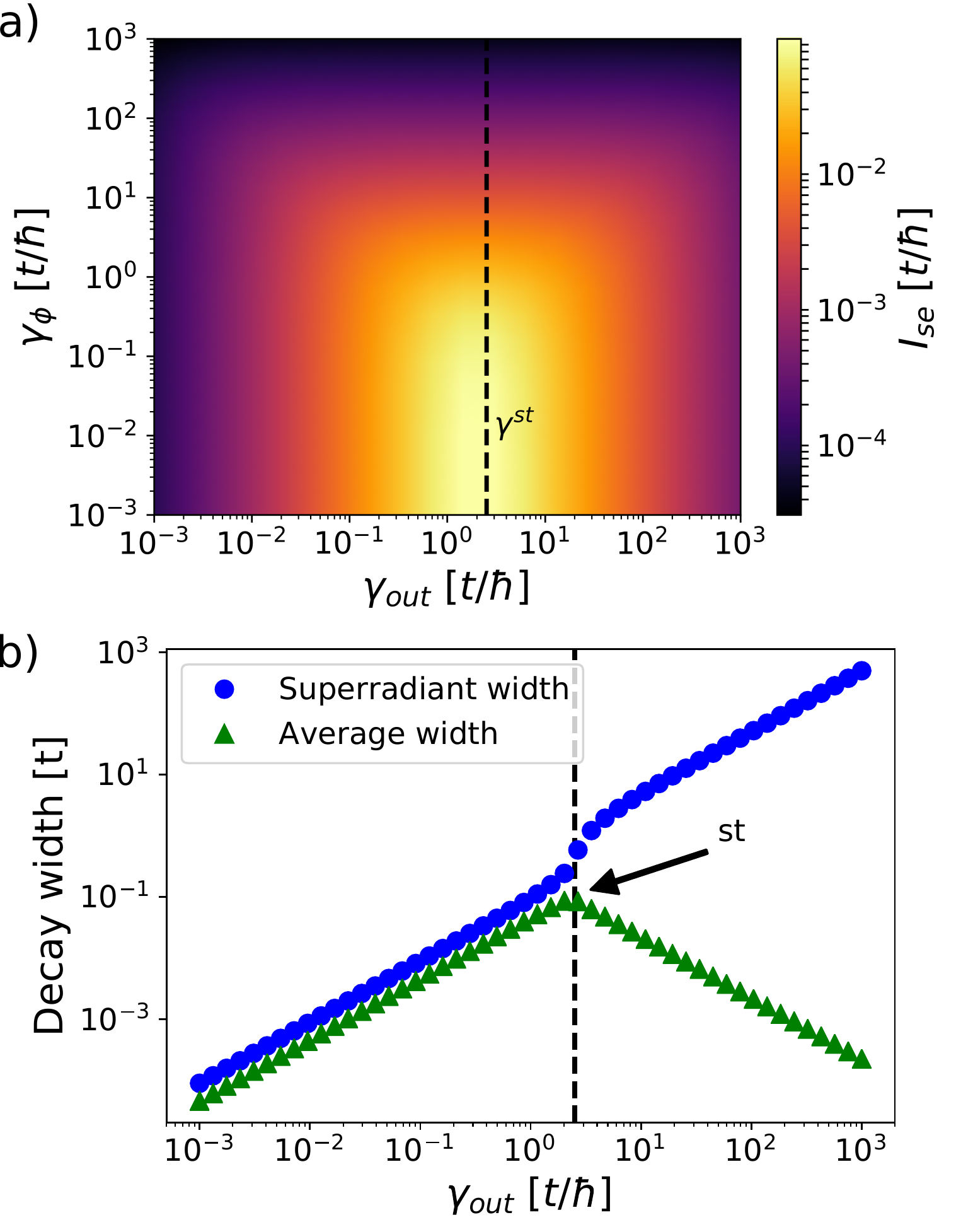}
	\caption{ [Color online]  Numerics for: (a) Maximal current, $I_\Imax$, obtainable in the single particle approximation from maximization of Eq.~\eqref{eq:current def} in a chain of length $N=10$, as a function of both the coupling to the sink $\gamma_\gout$ and dephasing $\gamma_\phi$. The vertical dashed line indicates the optimal coupling out which agrees with the superradiant transition $\gamma_\gout = \gamma^\st$ shown in the b) panel. (b) Decay widths as a function of the coupling to the sink $\gamma_\gout$ in absence of dephasing. Blue circles mark the largest width, green triangles indicate the average of all other widths. The dashed vertical line indicates the superradiant transition, namely where the average widths have a maximum.}\label{fig:OptimalCurrent}
\end{figure}

In Fig.~\ref{fig:OptimalCurrent}(a), the maximal current achievable within the single excitation approximation, $I_\Imax$, is shown as a function of the two parameters $\gamma_\gout$ and  $\gamma_\phi$ (for $\gamma_\grec = 0$). As one can see, for any fixed dephasing rate $\gamma_\phi$, the optimal value of $\gamma_\gout$ (for which the current is maximal as we will discuss in more details below) occurs at  the superradiant transition (st) (see Ref. \citenum{Celardo2009}), i.e., when $\gamma_\gout\simeq\gamma^{st}$. This is quite remarkable, and is in contrast with previous results obtained in the strong disorder regime where the presence of dephasing can strongly modify the value of the optimal coupling \cite{Zhang2017a}.

 The optimality at the superradiant transition can be physically explained as follows: the coupling to the sink induces a level broadening, namely each eigenstate of the chain acquires a finite width which represents a decay probability to the continuum as shown in Fig.~\ref{fig:OptimalCurrent}(b) (see also Appendix \ref{app:Superradiant}). For small $\gamma_\gout$ the decay widths increase linearly with the coupling to the sink and so does the decay probability. Hence, the current also increases with $\gamma_\gout$. When the average width becomes of the order of the average level spacing (overlapping of all resonances), one state becomes superradiant, namely it continues to increase its width as $\gamma_\gout$ is increasing, while all the others become subradiant, namely they start to decrease their widths as $\gamma_\gout$ is increasing, as shown in Fig.~\ref{fig:OptimalCurrent}(b). At this point it is no more convenient to increase further the coupling $\gamma_\gout$, since the vast majority of states involved in the dynamics under this action will now decrease their decay width. Therefore a maximal current will appear for $\gamma_\gout = \gamma^\textrm{st}$.

In Fig.~\ref{fig:OptimalCurrent}(b) the largest width (superradiant state) and the average width of all other states (subradiant) are shown as a function of $\gamma_\gout$ in absence of dephasing and losses. The average width of the subradiant states has a maximum at $\gamma^\st \sim 2t/\hbar$, indicated by a dashed vertical line in both Figs.~\ref{fig:OptimalCurrent}(a) and ~\ref{fig:OptimalCurrent}(b). Interestingly, this very simple estimate for the optimal $\gamma_\gout$ works even in presence of dephasing. This happens because we are dealing with a clean system (absence of static disorder and therefore of Anderson localization) and so the dephasing is always, independently on the value of $\gamma_\gout$, detrimental to transport, since it only induces a transition from ballistic to diffusive motion.

In the presence of static disorder, one expects the so-called dephasing assisted transport, in agreement with many results in literature (see, e.g., Refs. \citenum{Plenio2008,Mohseni2008}), for which an optimal value of the dephasing is found to be beneficial to transport.
However, in view of the application of our model to transport in transition metal oxides (c.f. Section \ref{sec:LVO}) which can be fabricated with a very low degree of disorder, we will ignore the effects of disorder in the following treatment. For a more detailed discussion about disorder see Appendix \ref{app:Disorder}.

In order to address the role of the different parameters and compare the optimal condition for transport to realistic values for TMO heterostructures we seek in the following an effective analytical formula to characterize the NESS current. Interestingly, we can derive such a formula 
using the phenomenological expression obtained in Ref.~\citenum{Zhang2017a} for a one-dimensional chain in presence of both static disorder and strong dephasing.

The starting point of our derivation is Eq.~(46) from Ref.~\citenum{Zhang2017a}, which describes the average transfer time when the initial excitation is concentrated in the first site, i.e., with $\rho(t_0) = \ketbra{1}{1}$ the density matrix at the initial time, as:
\begin{align}
	\tau &= \frac{N}{\gamma_\gout} + \frac{(N-1)(N-2)}{2 \Gamma_\textrm{F}} + \frac{N-1}{\Gamma_\textrm{L}} \nonumber \\&= \frac{N}{\gamma_\gout}+\frac{\hbar^2(N-1)}{4t^2}\left(N\gamma_\phi + \gamma_\gout\right) 	\label{eq:avg-transfer-time},
\end{align}
where
\begin{align}
	\Gamma_\textrm{F} &= \frac{2t^2}{\hbar^2\gamma_\phi} \label{eq:FoestersRate}\\
	\Gamma_\textrm{L} &= \frac{2t^2}{\hbar^2(\gamma_\phi + \frac{\gamma_\gout}{2})}. \label{eq:LeegwaterRate}
\end{align}
This simple expression has a very clear physical meaning. Eqs.~\eqref{eq:FoestersRate}, \eqref{eq:LeegwaterRate} define, respectively, the Föster \cite{Foerster1948} and the Leegwater \cite{Leegwater1996} transfer rates. While the former was introduced in order to describe the  energy transfer rate that characterizes the classical hopping in the limit of large dephasing, the latter is a phenomenological correction that takes into account the quantum coherent effects coming from the coupling to the sink. 

In order to derive the above equation, Zhang {\it et al.}~\cite{Zhang2017a} treated the average time as the result of different contributions. One has to add the transfer time  from the last site to the sink, $N/\gamma_\gout$, the diffusion time from the first to the second last site, $(N-1)(N-2)/(2\Gamma_\textrm{F})$, and the transfer time from the second last to the last site, $(N-1)/\Gamma_\textrm{L}$. This is clearly understandable in the limit of large dephasing where the dynamics in the chain is expected to be diffusive. Quite surprisingly, Eq.~\eqref{eq:avg-transfer-time} turns out to be valid even for $\gamma_\phi = 0$ when the transport through the chain becomes ballistic (and not diffusive as the simple argument above was suggesting), as indicated by the comparison with our numerical simulations in Fig.~\ref{fig:Analytics}(a). This can be understood considering that, for short chains, the average transfer time from ballistic propagation is negligible with respect to the time-scale set by the sink. \ch{Note that Eq.~\eqref{eq:avg-transfer-time} was derived heuristically in the semi-classical limit, and not from first principles. 

\begin{figure*}
	\centering
	\includegraphics[width = 0.9\textwidth]{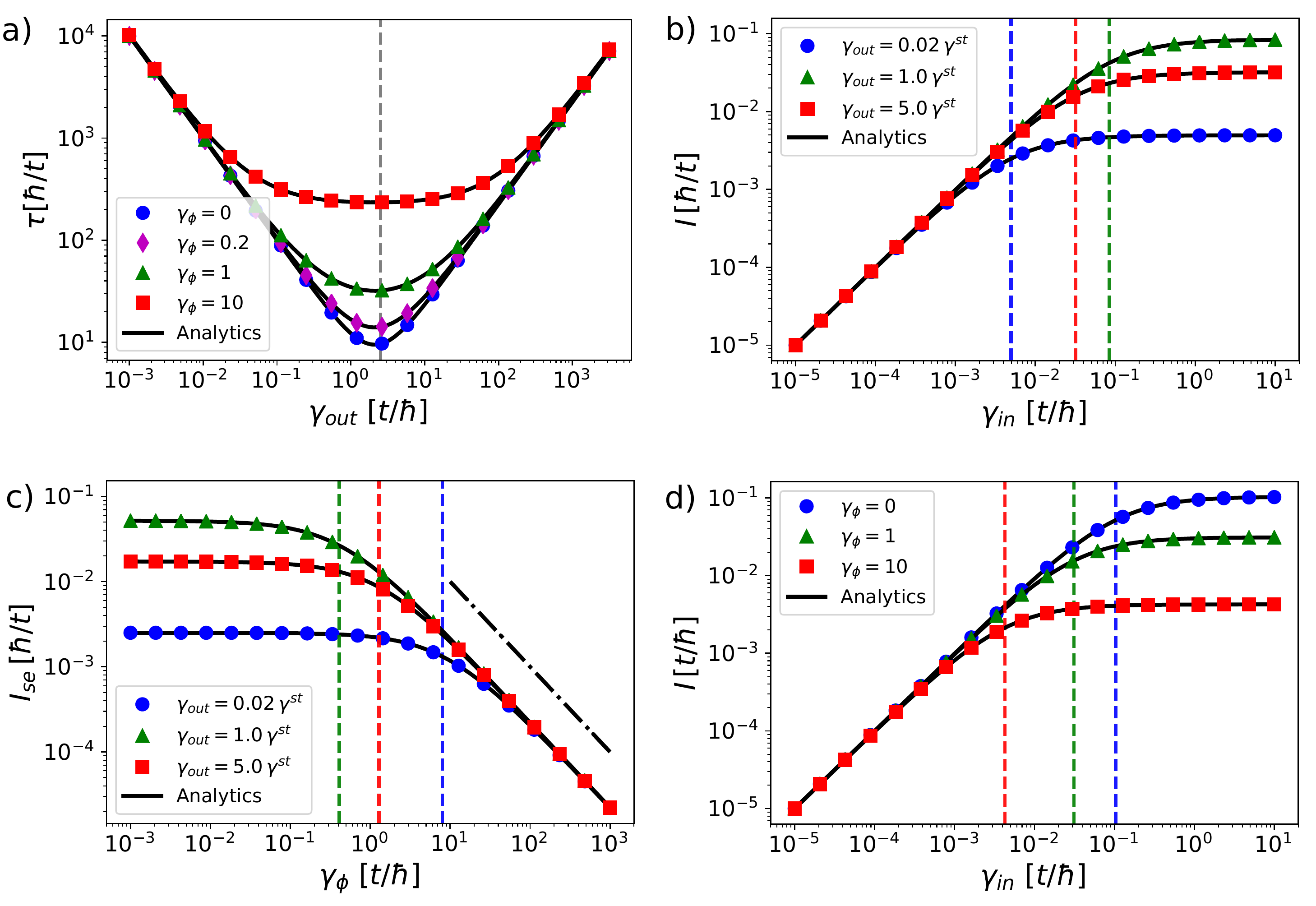}
 	\caption{[Color online] Analytical formula (black solid line) and numerics (markers) for: (a) average transfer time, Eq.~\eqref{eq:avg-transfer-time}, (b,d) NESS current, Eq.~\eqref{eq:CurrentAnalytical}, and (c) maximal current in the single particle approximation, Eq.~\eqref{eq:imax}. (a) The average transfer time is minimal at the superradiant transition $\gamma^\st$ (vertical dashed line) and increases with increasing values of $\gamma_\phi$. At large $\gamma_\phi$, the minimum peak turns into a plateau. Far from the superradiant transition, $\tau$ is almost independent of $\gamma_\phi$. (b,d) Below $1/\tau$ (vertical dashed lines) the current is linearly proportional to $\gamma_\gin$. \ch{Above $1/\tau$ the current saturates because all the calculations are here done in the single excitation subspace (c.f. Appendix \ref{app:Many-particles}).}  For (b) $\gamma_\phi =0.1 t/\hbar$ and for (d) $\gamma_\gout = \gamma^\st$. (c) Below the value of the dephasing $\crit{\gamma}_\phi$, Eq.~\eqref{eq:MaxDephasing1}, (vertical dashed lines) the current $I$ is independent of $\gamma_\phi$. Above $\crit{\gamma}_\phi$ the current decreases $\propto \gamma_\phi^{-1}$ (dash-dotted line).}
\label{fig:Analytics}
\end{figure*}

It is thus quite remarkable that an effective analytical formula for the NESS current that agrees perfectly with our numerical simulations even in the quantum regime (see Figs.~\ref{fig:Analytics}(b-d)) can be derived from Eq.~\eqref{eq:avg-transfer-time}, in the single excitation approximation ($\gamma_\gin < 1/\tau$, see also Appendix \ref{app:Many-particles}) when both the disorder and the losses are small (c.f. Appendix \ref{app:Disorder})}. To begin with, we note that in the single excitation approximation, the maximum achievable current $I_\Imax$ is obtained when the rate of incoming excitations $\gamma_\gin$ is equal to the inverse of the average transfer time $1/\tau$. In other words, a new excitation can enter the system only when the previous one has escaped. For pumping rates larger than the average transfer time ($\gamma_\gin \gg 1/\tau$), the single excitation approximation breaks down (see Appendix \ref{app:Many-particles}). The \ch{limit of the single excitation approximation can be observed in Fig.~\ref{fig:Analytics}(b,d) when the current saturates for $\gamma_\gin > 1/\tau$. At this point, the system could transmit more than one excitation at a time, but since we limit ourselves to the single excitation approximation, the current saturates}. At pumping rates smaller than the saturation value, the current is directly proportional to the pumping rate. This appears natural, since all the excitations that enter the chain leave the chain in a time that is shorter than the inverse pumping rate. Since there are no losses in the chain, all injected excitation leave the system. Thus, we can express the NESS current as
\begin{align}\label{eq:CurrentAnalytical}
	I = \frac{\gamma_\gin}{1 + \gamma_\gin\tau}.
\end{align}
Remarkably, our heuristic formula for $I$ perfectly agrees with our numerical simulations of the stationary current, given that there is no disorder nor spontaneous emission in the chain as illustrated in Fig.~\ref{fig:Analytics}(b,d). Hence, we can derive the maximum possible current in the single excitation approximation by setting $\gamma_\gin = \tau^{-1}$ in Eq.~\eqref{eq:CurrentAnalytical},
\begin{align}\label{eq:ImaxSingle}
	I_\Imax := \frac{1}{2\tau}.
\end{align}
 
Using the analytical expressions reported in Eqs.~\eqref{eq:avg-transfer-time},\eqref{eq:CurrentAnalytical}, and \eqref{eq:ImaxSingle}, the key parameters characterizing the NESS current and its maximum value can be derived. We remark that maximizing the NESS current $I_\Imax$ is equivalent to minimizing the average transfer time \eqref{eq:TauDef}. Moreover, the current \eqref{eq:CurrentAnalytical} does depend on the system's parameter only through the average transfer time $\tau$. Hence, we will focus in the following discussion on current optimization, on the role of $\tau$.

The average transfer time exhibits a minimum as a function of the value of the coupling to the sink which can be derived from Eq.~\eqref{eq:avg-transfer-time} as
\begin{align}
	\gamma_\gout^\textrm{opt} = \frac{2 t}{\hbar} \sqrt{\frac{N}{N-1}} \approx  \gamma^\st \simeq \frac{2t}{\hbar} . \label{eq:OptimalOpening}
\end{align}
As it was argued in Ref.~\cite{Celardo2012}, this minimum is in good agreement with the superradiant transition value $\gamma^\st$. The cooperative quantum coherent nature of this effect \cite{Sokolov1992} suggests that the current is optimized due to cooperativity and quantum coherences for $\gamma_\gout \approx \gamma^\st$. Such an interpretation is confirmed by the fact that, in the semiclassical regime, i.e., when dephasing becomes dominant, $\gamma_\phi \gg \gamma_\gout$, the minimum of the average transfer time as a function of $\gamma_\gout$ is broadened as shown in Fig.~\ref{fig:Analytics}(a). Moreover, we note that for large dephasing, the Leegwater rate \eqref{eq:LeegwaterRate} converges to the Föster rate \eqref{eq:FoestersRate}, $\Gamma_L \approx \Gamma_F$. Hence, according to Eq.~\eqref{eq:avg-transfer-time}, the transport is then driven by pure incoherent hopping only. Therefore, we can define a value of the dephasing $\crit{\gamma}_\phi$ above which the coherent effects leading to a minimization of the average transfer time are not relevant, and above which there is thus no real gain in working at the superradiant transition $\gamma_\gout^\textrm{opt} \approx \gamma^\st$. The dephasing rate can be estimated from Eq.~\eqref{eq:avg-transfer-time} by comparing the leading order in $N$ components of $\gamma_\phi$ and $\gamma_\gout$:
\begin{align}
	\crit{\gamma}_\phi(\gamma_\gout,N) & \sim \frac{\gamma_\gout^2 + 4 t^2 / \hbar^2}{N\gamma_\gout}. \label{eq:MaxDephasing1}
\end{align}
In Fig.~\ref{fig:Analytics}(c) we see that for any value of $\gamma_\gout$, when $\gamma_\phi \leq \crit{\gamma}_\phi$ the average transfer time is largely independent on $\gamma_\phi$. On the contrary, for  $\gamma_\phi \geq \crit{\gamma}_\phi$, the current is decreasing as $1/\gamma_\phi$. We conclude that in order to take full advantage of the superradiant transition for maximizing the current, we require that $\gamma_\phi \leq \crit{\gamma}_\phi$. More precisely, this maximum allowable dephasing, \eqref{eq:MaxDephasing1}, at $\gamma_\gout = \gamma_\gout^\textrm{opt} \approx \gamma^\st$, \eqref{eq:OptimalOpening}, is given by  
\begin{align}\label{eq:CriticalDephasing}
	\crit{\gamma}_\phi(N) = \frac{2t}{\hbar}\frac{2N-1}{N\sqrt{N(N-1)}} {\simeq} \frac{4t}{\hbar N}
\end{align}
As $\crit{\gamma}_\phi$ in Eq.~\eqref{eq:CriticalDephasing} scales gently with the chain's length N, the coherence-enhanced current can be expected in short, albeit not experimentally impractical short, chains as we shall discuss in Section \ref{sec:LVO}.

\begin{table}
\center
\begin{tabular}{@{}lcl@{}}
\toprule
\textbf{Parameters} & ~& \textbf{Value} \\ 
\midrule
Single particle approximation & ~& $\gamma_\gin \leq \tau^{-1}$ \\ 
Optimal rate out &~& $\gamma_\gout^\textrm{opt} \simeq \frac{2t}{\hbar}\sqrt{\frac{N}{N-1}}$ \\ 
Max. dephasing &~& $\crit{\gamma}_\phi \simeq 4t(\hbar N)^{-1}$\\[4pt]
Max. disorder &~& $\crit{W} \simeq t$ \\ 
Max. loss rate &~& $\crit{\gamma}_\grec \simeq \tau^{-1}$ \\ 
Optimal current below max. values  & \quad\quad& $\crit{I}_\Imax > t(\hbar N)^{-1}$ \\
\bottomrule
\end{tabular} 

\caption{Summary of the key parameters of the model \eqref{eq:tight-binding equation}-\eqref{eq:LopRec}. The max. values indicate the level of external noises below which the current is optimized at the superradiant transition, i.e., $\gamma_\gout = \gamma^\st$. For the max. disorder and max. loss see Appendix \ref{app:Disorder}}.
\label{tab:Parameters}

\end{table}

At the same time, the maximal current obtained by saturating the single particle approximation bound (c.f., Eqs.~\eqref{eq:avg-transfer-time}\eqref{eq:ImaxSingle}) and using $\gamma_\gout = \gamma_\gout^\textrm{opt.}$ is
\begin{align}
\label{eq:imax}
	&I_\Imax(\gamma_\phi, N) = \Bigg( \frac{\hbar\sqrt{N(N-1)}}{t} +\frac{\hbar^2N(N-1)\gamma_\phi}{4t^2} \Bigg)^{-1}.
\end{align}
From Eq.~\eqref{eq:imax} it is clear that for a dephasing $\gamma_\phi \simeq \crit{\gamma}_\phi (N)$, \eqref{eq:CriticalDephasing}, the maximal current at $\gamma_\gout = \gamma_\gout^\st$ scales inversely proportional to the chain's length:
\begin{align}
	&\crit{I}_\Imax (N) > \frac{t}{\hbar\sqrt{N(N-1)}}\propto \frac{t}{\hbar N}. 
\end{align} 
This again indicates that we indeed have coherence enhanced current, since a classical current driven by incoherent hopping is expected to be diffusive-like, i.e., $\propto N^{-2}$. 
 
In brief, we established that transport can be made efficient by exploiting coherences and derived an analytical bound for the dephasing rate below which this can be done. So far, we neglected static disorder and losses in the chain. 
As we show in Appendix \ref{app:Disorder}, our coherence-enhanced current is robust 
against disorder before localization effects play an important role, i.e, for $W \ll \crit{W} \simeq t $, and is robust against losses for loss rates smaller than the inverse average transfer time $\gamma_\grec < \tau^{-1}$. These constraints are compatible with the estimated values for the TMOs heterostructures considered in Section \ref{sec:LVO}.

In summary, in our transport model \eqref{eq:tight-binding equation}-\eqref{eq:LopRec}, coherences can be exploited to optimize current because of two factors: first, (coherent) ballistic transport quickly drives the excitation to the collector and second, the excitation is very rapidly absorbed via the coupling with the short-lived superradiant solution, which is spatially localized at the interface with the collector. While the first mechanism is suppressed by dephasing, which drives the transport through the chain from ballistic to diffusive (in the limit of infinitely long chains), the second remains unaffected by it. However, the speed-up from a quick absorption is mitigated by the longer transport time to the collector (see Fig.~\eqref{fig:Analytics} a). The table \ref{tab:Parameters} summarizes the key parameter constraints and optimal values of our model.


\section{Strong correlation effects: Dynamical Mean-Field Theory}
\label{sec:DMFT}
Many properties of TMOs are determined by the strong repulsive interaction 
between electrons occupying the same atomic orbitals, 
which eventually leads to the failure of the single-electron approximation 
and the emergence of Mott-insulating states. As a further step towards the exploitation in real devices of the coherence-enhanced phenomena discussed in the previous section, we should address the role of the electronic interactions which underlie the Mott insulating ground state of the TMOs discussed in Sec.~\ref{sec:LVO}. The goal here is not however to study the non-equilibrium dynamics of correlated TMOs following a photoexcitation, but rather to introduce the simplest description of a correlated excitation and to compare it with the results obtained in Section \ref{sec:master_eq}. Throughout this section we will assume that we can describe the photoinjected carriers in terms of an effective photodoping of the Mott insulating ground state. It is well established by dynamical mean-field theory (DMFT) that the doping of a Mott insulator results in a narrow quasiparticle peak flanking the upper Hubbard band\cite{Georges1996} (assuming electron doping). For a small number of doped carriers, which is certainly the case we are interested in, the quasiparticle peak remains well separated from the Hubbard band. Nonetheless, the states in the peak have a small effective hopping and a finite lifetime due to the interaction with the other particles. 

In this section we study if the optimization of the quantum  transport at the superradiant transition survives in the presence of a strongly interacting environment by computing the linear response of the system as a function of the interaction strength. In the same spirit, we make one further approximation. For small finite system, the minimum doping which can be introduced to half-filling (N electrons on N sites) is set by the number of sites. Indeed, doping either adds or removes one electron, so the average charge per site is modified by $1/N$. This means that with doping we have to consider densities which are quite distant from the condition of half-filling, that thus yield a physical behavior very different from the one of a doped Mott insulator which we are interested in. For this reason we decided to model our system as a half-filled system with a Coulomb repulsion smaller than the critical value for the Mott transition, instead of a doped system. In this way we obtain a picture which includes the important features of the photoexcitations we are interested in.  

We performed equilibrium transport calculations of the tight-binding model of the effective chain in the presence of a local Coulomb interaction, denoted by an effective Hubbard parameter $U_{eff}$ (see the many-particle model illustration in Fig.~\ref{fig:hetero}), which measures the residual interaction involving the photoexcited carriers:
\begin{equation}\label{eq:Hubbard}
 H = t \sum_{j=1}^{N-1} c^{\dagger}_{j\sigma} c^{\phantom{\dagger}}_{j+1\sigma} 
   + \textrm{h.c.} 
   + U_{eff} \sum_{j=1}^{N} c^{\dagger}_{j\uparrow}  c^{\phantom{\dagger}}_{j\uparrow}
                      c^{\dagger}_{j\downarrow}c^{\phantom{\dagger}}_{j\downarrow},
\end{equation}
where $c^{(\dagger)}_{j\sigma}$ are the fermionic annihilation (creation) operators for an electron at site $j$ with spin $\sigma\!=\!\{\uparrow,\downarrow\}$. In the non-interacting limit $U_{eff}=0$, the Hubbard Hamiltonian~(\ref{eq:Hubbard}) is identical to the tight-binding model in Eq.~(\ref{eq:hamiltonian}) in the absence of disorder (and by setting $\epsilon_0=0$). By tuning the interactions, one can explore the effects of the electron-electron interaction on the superradiant transition. The electronic conductance is evaluated in the linear response regime. The conductance $g\!=\!(e^2/h) T(0)$ is defined  in terms of the conductance quantum ($e^2/h$), and the transmission function 
\begin{equation}
T(\hbar\omega)=\textrm{Tr}\Big[ \hat{\Gamma}_L \hat{G}^a(\hbar\omega) \hat{\Gamma}_R \hat{G}^r(\hbar\omega)\Big].
\end{equation}
The Green's function of the chain is defined given by
\begin{align}\label{eq:Hubbard-Greens}
 \hspace{-10pt}\hat{G}(\hbar\omega)= \Big[ \hbar\omega\hat{\mathbb{1}} - \hat{H_0} 
                             -\hat{\Sigma}_L(\hbar\omega)-\hat{\Sigma}_R(\hbar\omega)
                             -\hat{\Sigma}(\hbar\omega)
                      \Big]^{-1},
\end{align}
where $H_0$ is the tight-binding Hamiltonian of the effective chain model,  
$\hat{\Sigma}_{\alpha}$ is the self-energy of the left (L) and right (R) leads, 
and $\hat{\Sigma}$ accounts for the electronic correlations in the chain. 

The retarded (advanced) Green's function $\hat{G}^{r(a)}\!=\!\hat{G}(\hbar\omega \pm \imath\eta)$ describes the propagation of electrons (holes). The broadening $\eta$ is necessary for the regularization of the analytic continuation, and at the same time, it effectively accounts for contributions not included in the Hubbard model at zero temperature (e.g., electron-phonon coupling, thermal excitations, and other decay channels). It has the functional form of a local dissipative term, and is equivalent to the loss rate $\gamma_\textrm{loss} = 2\eta/t$ introduced in Eq.~(\ref{eq:LopRec}) in the quantum master equation formalism for the dephasing.

The matrix $\hat{\Gamma}_{\alpha}=i\big(\hat{\Sigma}^{r}_{\alpha}-\hat{\Sigma}^{a}_{\alpha}\big)$ encodes the spectral information of the leads. In the following we assume a wide-band limit (WBL) approximation for the sinks, so that the leads' self-energy takes the form $\hat{\Sigma}_{L}=-i\gamma|1 \rangle\langle 1|$ and $\hat{\Sigma}_{R}=-i\gamma|N \rangle\langle N|$. The presence of $\hat{\Sigma}_{L/R}$ introduces another source of broadening of the Hamiltonian eigenstates, which is equivalent to the one shown in Fig.~\ref{fig:OptimalCurrent}(b), where $\gamma / t$ here plays the role of $\gamma_{\gout}$. The main difference with the Lindblad master equation approach is that the system is not open, and despite the coupling to the sink, the electron number is conserved (and fixed to one electron per site). In this configuration, the transmission 
\begin{equation} \label{eq:transmission}
 T(\hbar\omega) = \gamma^2 |G^{r}_{1N}(\hbar\omega)|^2
\end{equation}
is the probability of the electron to be transmitted from one end to the opposite end of the chain.

Electronic correlations are taken into account through the self-energy matrix $\hat{\Sigma}(\hbar\omega)$. The self-energy is obtained within the framework of real-space dynamical mean-field theory (DMFT),\cite{Snoek2008} which allows to obtain the electronic\cite{Valli2016,Valli2015} and transport properties\cite{Valli2010,Jacob2010,Valli2010,Mazza2016,Valli2018,Valli2018a} of models and realistic Hamiltonians for correlated systems which lack translational invariance in one or more spatial dimensions.\cite{Schuler2017} Within real-space DMFT, the self-energy fully retains quantum fluctuations, but it is approximated to be local, yet site-dependent, i.e., $\Sigma_{ij}(\hbar\omega)=\Sigma_{ii}(\hbar\omega) \delta_{ij}$, so that non-local spatial correlation are retained just at the static mean-field level. 
The imaginary part of the self-energy correspond to an energy-dependent broadening of the energy levels, while the real part accounts for a renormalization of the energy of the quasi-particles (i.e., the poles of the Green's function). The dynamical character of $\hat{\Sigma}(\hbar\omega)$ is fundamental to describe strong correlations effects, including (but not limited to) the Mott charge localization due to the interactions. 
While single-site DMFT is known to have issues 
in one dimension,~\cite{Capone2004} extensions such as 
two-site cellular DMFT and real-space DMFT has been shown to provide 
quantitative and/or qualitative agreement 
with (numerically) exact methods such e.g., 
density-matrix renormalization group.~\cite{Capone2004,Wernsdorfer2011,Schwabe2013}

The auxiliary impurity models for each site in the chain are solved at $T=0$ with an exact diagonalization solver~\cite{Capone2007} with $n_b=8$ bath sites. 
The knowledge of the impurity spectrum allows to evaluate 
the retarded self-energy directly, 
avoiding ill-defined analytic continuation procedures, 
and it is thus suitable to describe 
the transport~\cite{Valli2018,Valli2018a}
Since the Green's function is dressed by the DMFT self-energy as in Eq.~(\ref{eq:Hubbard-Greens}), the transmission probability~(\ref{eq:transmission}) fully takes into account all scattering processes within the chain described by  $\hat{\Sigma}(\hbar\omega)$. 

\begin{figure*}
	\includegraphics[width = 0.99\textwidth]{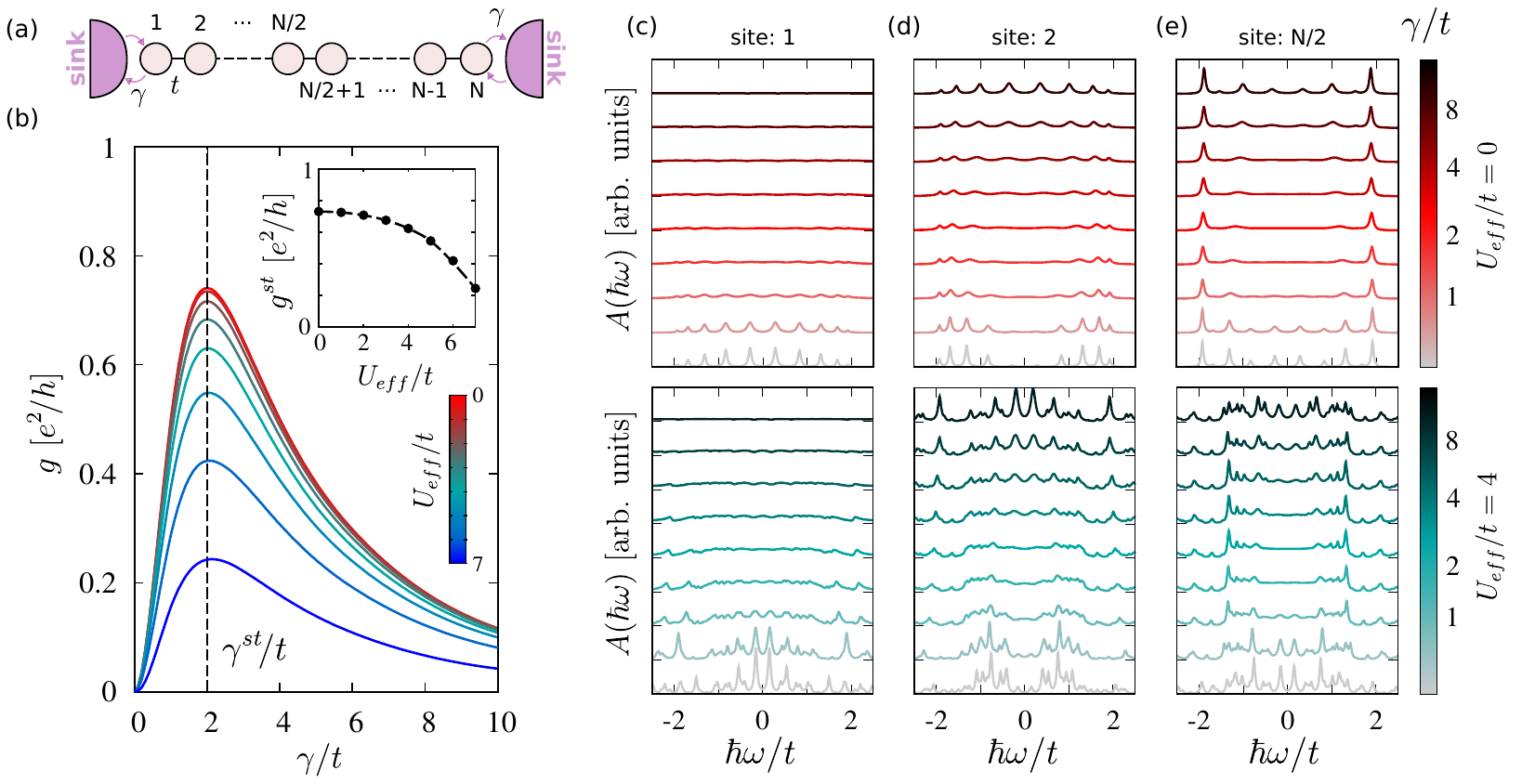}
	\caption{[Color online] (a) Sketch of the one-dimensional model solved within DMFT. 
    (b) Conductance $g$ (in units of the conductance quantum $e^2/h$) 
    as a function of the coupling to the sinks $\gamma$ 
    for different values of the local Hubbard interaction $U_{eff}$.
    The maximum of $g$, marker of the superradiant transition, 
    is found at approximatively $\gamma^{\st}\!=\!2t$ (dashed black line) 
    and it is \emph{nearly independent} on $U_{eff}$.
    Inset: suppression of the conductance at the superradiant transition $g^{st}$ 
    as a function of $U_{eff}$. 
    (c) Evolution with $\gamma$ of the site-resolved local spectral function 
    $A(\hbar\omega)$, for the sites labeled as in panel (a). 
    The states localized at the edge of the chain become superradiant 
    both at $U_{eff}\!=\!0$ and at finite $U_{eff}$. 
    The effect is symmetric on the other half of the chain.
    The results are obtained for $N\!=\!10$ with a regularization $\eta/t=0.03$ 
    for the analytic continuation of the retarded Green's function.}
	\label{fig:DMFT}
\end{figure*}

Let us first discuss the transport properties for the case $U_{eff}=0$. In this case, it can be shown that the conductance $g$ is related to the NESS current as $I \approx t/(N \hbar)\sqrt{g h/e^2}$ for $\gamma_\textrm{loss} = 2\eta /t$, $\gamma_\gout = \gamma/t$, $\gamma_\phi = 0$, $\eta \ll \gamma$ and $\gamma_{in} \gg  \gamma_\textrm{loss}$. Hence, the Lindblad master equation approach and the linear response transport within DMFT indeed describe two sides of the same coin. 

In Fig.~\ref{fig:DMFT}(b) we show the linear response conductance as a function of the coupling to the sinks $\gamma$  for different values of the local Hubbard interaction $U_{eff}$.  Remarkably, $g$ displays a maximum consistent with a superradiant transition 
at $\gamma^{\st}\!\approx\!2t/\hbar$, and this even for an interaction $U_{eff} \gg t$. 
Note that here the value of the superradiant transition $\gamma^\st$ 
is the same for the Hubbard model and the quantum master equation model.
The position of the maximum is \emph{nearly independent} on the value of $U_{eff}$. 
The main effect of electronic correlations is to \emph{suppress} the conductance, as shown in Fig.~\ref{fig:DMFT}(b) (see also
\cite{Oguri1997,Oguri1999,Oguri2000,Valli2012}),
which suggests a tempting analogy between the effect of electronic correlations on the conductance
and the effect of dephasing on the NESS current $I_\textrm{se}$ (c.f. Fig.~\ref{fig:OptimalCurrent}(a)). 
Moreover, the suppression of the conductance is minimal at the superradiant transition, 
demonstrating that the realization of the optimal conditions 
for coherent transport is of pivotal importance in the presence of strong correlations. 

In Fig.~\ref{fig:DMFT}(c-e) we visualize  the site-resolved local spectral function $A(\hbar\omega)=-\Im G^r(\hbar\omega)/\pi$ as a function of $\gamma/t$ across the superradiant transition.
Let us consider the non-interacting case $U_{eff}/t\!=\!0$ (upper panels) first. 
Below the optimal value $\gamma^{\st}\approx 2t$, the lifetime of the tight-binding eigenstates 
decreases upon increasing $\gamma$, 
and, correspondingly, the spectral features become broader. 
At the superradiant transition we observe that the states localized 
at the edge of the chain become superradiant, 
i.e., their widths become much larger than the widths of the central states, 
in complete analogy with the results of Fig.~\ref{fig:OptimalCurrent}(b)
and the tight-binding eigenvalue analysis done in Appendix \ref{app:Superradiant}. 
Remarkably, this scenario holds also in the presence of interactions, 
as shown for the case $U_{eff}/t\!=\!4$ (lower panels). 
For $\gamma>\gamma^\st$, the superradiant state is localized at the edge of the chain, 
while the excitation spectra of the inner sites
evolve into a collection of sharp resonances which persist up to $\gamma\gg t$. 

The Hubbard interaction discussed in this context provides 
a simplified description of the interactions in TMOs, 
as it lacks the intrinsic multi-orbital nature of the electronic correlations, 
as well as the interplay between charge, spin, and lattice degrees of freedom. 
At the same time, this minimal model represents the ideal framework 
to discuss the role of electronic correlations on the superradiant transport 
beyond the Lindblad master equation approach. 
In particular, it allows to make an analogy between the effect 
of dephasing ($\gamma_{\phi}$) and electron-electron interactions  
as described by the Hubbard $U_{eff}$ term. 
Remarkably, the model also shows that the enhancement of the transport properties
due to coherent-assisted effects survives up to moderate interaction strength, 
which falls in the relevant range for TMOs. 
This suggests that the physics of the superradiant transition 
is relevant (and can be straightforwardly extended) to electric and thermal transport 
in quantum junctions. 
In the literature, the signature of superradiant electron transport 
through a quantum dot was discussed in terms of the coherent dynamics 
induced by hyperfine interaction \cite{Schuetz2012}, 
and it was recently shown that the superradiant effect 
in thermal emitters leads to an abnormal power scaling, which could be exploited 
to engineering efficient energy conversion devices \cite{Zhou2015}.


\section{Outlook for an application to heterostructures}
\label{sec:LVO}
The observation and exploitation of coherence-enhanced transport in the vicinity of the superradiant transition builds on the possibility of collecting charges in real devices on timescales faster than the decoherence time. Eq.~\eqref{eq:CriticalDephasing} defines the maximum number of sites of the chain which allows coherent transport, i.e. $N\simeq$4$t$/$\hbar\crit{\gamma}_{\phi}$. The large value of  $\crit{\gamma}_{\phi}$ for solid state systems at high temperatures therefore requires the design of devices composed by few atomic layers. TMOs heterostructures are one of the most promising platform to investigate the high-temperature coherent phenomena discussed in the present work. The chemical composition, the interface strain mismatch and disorder, and the number of atomic layers can be controlled to a level \cite{Hwang2012} that makes the design of nano-devices with coherence-driven functionalities a real opportunity. In this section we will consider the case of (SrVO$_3$)$_{n}$/(LaVO$_3$)$_{m}$/(SrVO$_3$)$_{n}$ heterostructures, $n$($m$) being the number of atomic layers, which provides an interesting example of a realistic system where the conditions discussed in Tab. \ref{tab:Parameters} can be fulfilled. 

\begin{figure*}
	\includegraphics[width = 1\textwidth]{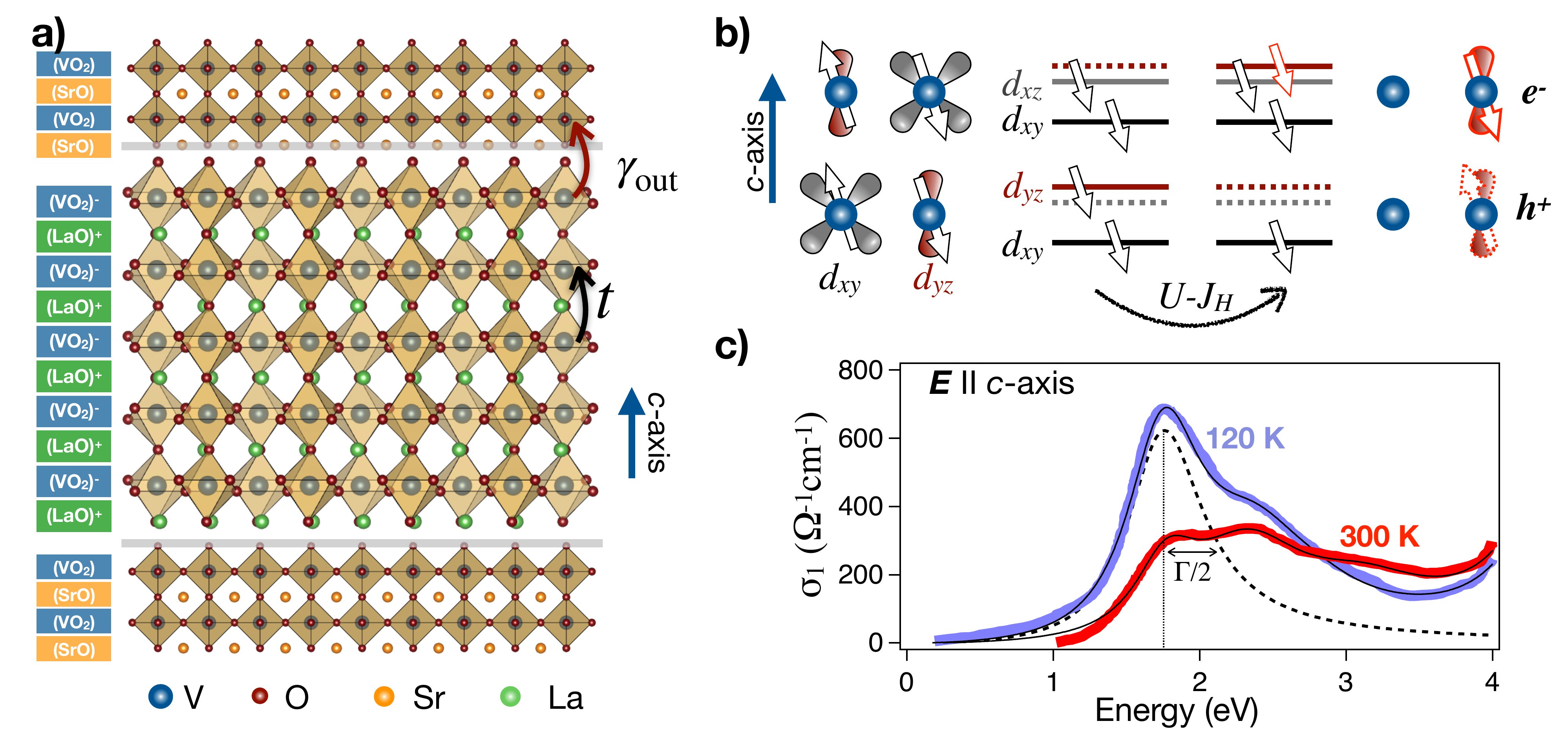}
	\caption{[Color online] a) Perovskite structures of LaVO$_3$ and SrVO$_3$. The few-monolayers (SrVO$_3$)$_{n}$/(LaVO$_3$)$_{m}$/(SrVO$_3$)$_{n}$ heterostructure constitutes a promising platform to realize coherence-enhancement transport schemes. b) Cartoon of the real-space electronic configuration (left and right panels) and energy-levels scheme (central panels) of the $d_{yz}$-$d_{yz}$ optical transition, which results in the photo-injection of an electron-hole excitation. The black arrows indicate the electronic spin configuration in the orbital- and spin-ordered state at $T<T_c$. The red arrows indicate the spin configuration of the photo-excitation. The energy cost for the $d_{yz}$-$d_{yz}$ optical transition is $U$-$J_H\simeq$1.8 eV, $J_H$ being the Hund's coupling. c) Real part of the $c$-axis optical conductivity of bulk LaVO$_3$, as measured by optical spectroscopy. The data are taken from Ref.~\cite{Miyasaka2002}. At $T<T_c$, the optical conductivity evidences the appearance of a well defined absorption feature at $\simeq$1.8 eV for light polarization parallel to the $c$-axis. The dashed line represents the Lorentz oscillator, which accounts for the 1.8 eV optical transition. The estimated transition width is $\Gamma$/2$\simeq$300 meV.}
	\label{fig:hetero2}
\end{figure*}

Bulk LaVO$_3$ (LVO) is a Mott insulator with +3 nominal valence of the V atoms. At room temperature, LVO crystallizes in an orthorombic structure, with lattice parameters $a$=5.555 {\AA}, $b$=5.553 {\AA}, and $c$=7.849 {\AA}. The structure of LVO (see Fig. \ref{fig:hetero}(a)) can be derived from the cubic perovskite by tilting the VO$_6$ octahedra in alternating directions around the $b$-axis and rotating them around the $c$-axis \cite{DeRaychaudhury2007}. The cubic crystal field given by the oxygen cage surrounding the V atoms splits the V-3$d$ levels into a doubly occupied $t_{2g}$ triplet ($xy$, $yz$, $xz$ symmetries) and an unoccupied $e_g$ doublet ($x^2$-$y^2$, 3$z^2$-$r^2$ symmetries) separated by a $\sim$2 eV gap \cite{Wang2015}. In contrast with band-theory predictions, the onsite Coulomb repulsion related to the double occupation of the $t_{2g}$ orbitals leads to the freezing of the charge carriers motion and the consequent formation of a Mott insulating ground state with a gap of the order of 1.1 eV \cite{DeRaychaudhury2007,Assmann2013,Wang2015}. The on-site Coulomb repulsion $U_\textrm{LVO}$ is estimated as 3-5 eV, when the same orbital is doubly occupied by two electrons with opposite spin, and 1-3 eV, when the double occupation involves two different  $t_{2g}$ orbitals with parallel spin configuration (see Fig. \ref{fig:hetero2}(b)), as favored by the Hund's coupling \cite{Miyasaka2002,DeRaychaudhury2007}. The oxygen mediated hopping integrals connecting vanadium atoms on different sites are strongly orbital-dependent. When considering the interlayer hopping, i.e. along the $c$-axis, the dominant terms are given by the $d_{xz}\rightarrow d_{xz}$ process with $t_{xz,xz}\simeq$200 meV ($\hbar$/$t_{xz,xz}\simeq$3.3 fs) and $d_{yz}\rightarrow d_{yz}$ with $t_{yz,yz}\simeq$130 meV ($\hbar$/$t_{yz,yz}\simeq$5 fs) \cite{DeRaychaudhury2007}.

As recently suggested \cite{Assmann2013,Wang2015}, LVO can be used as the main building block for the development of innovative photovoltaic devices. Besides the perfect matching between the LVO optical absorption (see Fig. \ref{fig:hetero}(c)) and the solar spectrum, a fundamental characteristic is the potential gradient along the (LaO)$^+$-(VO$_2$)$^-$ planes (see Fig. \ref{fig:hetero}(a)), which spontaneously forms when LVO is sandwiched between non-polar perovskites (e.g. SrTiO$_3$, SrVO$_3$). This intrinsic potential \cite{Assmann2013} can be properly combined with the application of an external electric field, thus favoring the separation of photo-induced electron-hole excitations and the fast migration of photo-generated charges along the $c$-axis, i.e., in the direction perpendicular to the (LaO)$^+$-(VO$_2$)$^-$ planes. The combination of these ingredients opens the possibility to test the coherent-transport model presented in the previous sections. 

If we neglect the direct electron-hole interaction, the absorption of a photon with energy larger than the optical gap of 1 eV generates a single excess electron (hole) in the upper (lower) Hubbard band which migrates towards the maximum (minimum) of the electric potential by hopping along the $c$-axis with a hopping term as large as 200 meV and experiencing an effective interaction  $U_{eff}$<$U_{LVO}$. 
In the coherent-transport model, the rate of creation of photo-excitations corresponds to the rate $\gamma_\gin$. We point out that realistic working conditions for thin LVO-based heterostructures falls well within the single-excitation approximation, i.e. when the rate of incoming excitations $\gamma_{in}$ is smaller than the inverse of the average transfer time 1/$\tau$. For example, considering that the typical sunlight power in the visible is 450 W/m$^2$ and assuming an average photon energy of 2.3 eV, we obtain a photon flux that is 1.2$\cdot$10$^{17}$ s$^{-1}$cm$^{-2}$. Considering that the penetration depth of the visible light in LVO is of the order of 100 nm, the number of photons absorbed in a 5 nm thick LVO layer is less than 5\%. The photon flux absorbed is thus about 6 fs$^{-1}$cm$^{-2}$. Even considering a transfer time as long as 100 fs, it turns out that the number of absorbed photons in the time span $\tau$ is 600 cm$^{-2}$, which represents a very small density as compared to the surface-projected density of the vanadium atoms. Despite the fact that the light absorption of a very thin LVO film is rather small, the LVO layer, sandwiched between two metallic contacts, should be considered as the main building block to engineer devices whose efficiency relies on coherence-enhanced phenomena. The light absorption can be easily tuned  by stacking several building blocks up to the point of achieving the desired absorption efficiency. Thus, in order to build an efficient light-harvesting device that can harvest the power of the coherent-absorption at the superradiant transition, several LVO layers sandwiched between conducting layers are stacked to absorb a large portion of the incident light \cite{Assmann2013}.

In order to describe the dynamics of the charge migration and collection in the heterostructure via the one-dimensional coherent-transport model, we will assume that all the in-plane interactions can be accounted for by the decoherence rate $\gamma_{\phi}$. It is thus crucial to provide a quantitative estimation of $\gamma_{\phi}$ in the realistic case of the photoexcited charge carriers hopping in a fluctuating lattice, spin and charge background. A natural source of decoherence is constituted by the interaction with the thermally-activated phonons, whose inverse energy provides a lower limit for electron-phonon scattering time. In LVO, the highest energy modes which are coupled to electronic excitations are related to the rotation of VO$_6$ octahedra (23 meV), to oxygen bending and Jahn-Teller modes (35, 53 meV), which sets the minimum timescale for electron-phonon scattering to $\hbar$/53 meV=12 fs.

The timescale related to the direct charge-charge interaction is of more difficult evaluation. At room temperature, the average occupation of the $d_{xy}$, $d_{xz}$, $d_{yz}$ orbitals is 2/3 \cite{DeRaychaudhury2007}. The strong quantum fluctuations of the orbital occupation are thus expected to lead to an extremely rapid decoherence, which would kill the possible onset of coherent-transport regime. 

A possible solution to this problem is connected with the structural, magnetic and orbital ordering phase transition at $T_c\simeq$140 K. Below $T_c$ the lattice is subject to a 3\% elongation of the V-O bonds, which leads to the deformation of the VO$_6$ octahedra and the appearance of a new Raman-active oxygen stretching mode at 89 meV \cite{Miyasaka2006}. The structural phase transition is accompanied by the onset of a long-range C-type antiferromagentic order (antiferromagnetic in the $ab$-plane and ferromagnetic along the $c$-axis), with in-plane exchange energy $J_{ab}$=2 meV and inter-plane exchange energy $J_{c}$=33 meV \cite{Motome2003}. Interestingly, in the low-temperature phase the orbital fluctuations are almost completely suppressed and the system exhibits an orbital order corresponding to alternate $d^1_{xy}d^1_{xz}$/$d^1_{xy}d^1_{yz}$ occupation along the $c$-axis\cite{DeRaychaudhury2007}, as schematically shown in Fig. \ref{fig:hetero2}(b). In the orbitally ordered phase, the photo-generated charge excitations can thus move along a well-defined conductive channel, which connects the $d_{yz}$-$d_{xz}$ empty orbitals with a hopping integral $t_{yz,xz}\simeq$150 meV. At the same time, the coupling to collective orbital excitations is limited to 65 meV, as inferred by Raman measurements \cite{Miyasaka2005}.
We can thus conclude that the $c$-axis motion of photo-generated charges in the low-$T$ orbital-ordered phase of LVO is characterized by a hopping term $t_{yz,xz}$, that is almost twice as large as the energy scale of the fastest interaction (phonon modes at 89 meV), which represents the leading contribution to the $\hbar \crit{\gamma}_{\phi}$ term.
In the following, this hopping term $t_{yz,xz}$ will therefore play the role of the hopping $t$ introduced in the tight-binding and Hubbard models of Section II and III.

As a further confirmation of this picture, in Fig.~\ref{fig:hetero2}(c) we report the LVO optical conductivity as measured in Ref.~\cite{Miyasaka2002} by equilibrium optical spectroscopy. A strong peak centered at $\approx$1.8 eV appears in correspondence of the orbital-ordering phase transition and dominates the LVO absorption at $T<T_c$. The linewidth of this transition, corresponding to the hopping of an electron from a $d_{yz}$($d_{xz}$) orbital to the empty $d_{yz}$($d_{xz}$) neighboring orbital, is $\Gamma/2\approx$0.3 eV, corresponding to a lower bound for the lifetime of $\approx$2 fs, which is of the order of the transfer time in a device made of few monolayers.  Using these values we can thus estimate the critical thickness for the observation of coherence-enhanced transport, i.e. $N\simeq$4$t_{yz,xz}$/$\hbar\crit{\gamma}_{\phi}\simeq$8, that is well within the current technological possibilities.

As a last step, we discuss the role of the interfaces as efficient collectors of the photo-generated charges. The state-of-the art techniques for materials synthesis allows to grow few-monolayers LVO films with a degree of disorder much smaller than $\crit{W}$ and whose interface can be optimized for transport purposes. Since the optimal transport condition is achieved in the vicinity of the \textit{superradiant} condition, i.e. $\hbar\gamma_\gout^\textrm{opt}\simeq 2t_{yz,xz}$, a proper engineering of the collector is crucial for our purposes. A very promising route is terminating the few-monolayers LVO device with SrVO$_3$ (SVO), which is a non-polar cubic perovskite with almost perfect lattice matching. The +4 nominal valence of the V atoms corresponds to a single occupation of the $d_{xy}$, $d_{xz}$, $d_{yz}$ orbitals and a consequent metallic behavior down to a thickness of two-monolayers \cite{Yoshimatsu2010}. The very good orbital overlap between LVO and SVO orbitals, combined with the strongly correlated nature of the SVO metallic state \cite{Zhang2016}, thus provide a very effective scheme to transfer the charges photogenerated in LVO across the Mott gap to the collectors, where the electronic interactions rapidly leads to the down-conversion of the high-energy excitations into low-energy charge carriers and then into a detectable electrical signal.


\section{Conclusions} 
In conclusion, (SrVO$_3$)$_{n}$/(LaVO$_3$)$_{m}$/(SrVO$_3$)$_{n}$ heterostructures with $n\!>\!2$ and $m\!<\!8$ layers constitute a very promising platform to practically realize \cite{Sheets2007,Jeong2011} the conditions necessary to obtain \textit{superradiant}-assisted quantum transport at temperatures as high as 140 K. Being aware of the many simplifications contained in the quantum-trasport model presented here, the main goal of this work is to provide a general framework to guide the search for quantum-driven phenomena in solid-state devices working at ambient conditions. The results presented here are expected to boost the development of first-principle calculations \cite{Janson2018} to account for the complexity of the real devices, such as the interfacial lattice and electronic reconstruction and the scaling of LVO and SVO bulk properties with the device size. From the experimental standpoint, this work will trigger the development of techniques \cite{Giannetti2016,Gandolfi2017} to directly measure the electronic decoherence dynamics in correlated heterostructures.


\begin{acknowledgments}
We thank G.G. Giusteri for helpful discussions. F.B. and C.M.K. acknowledge support by the Iniziativa Specifica INFN-DynSysMath. M.C. acknowledges support from  H2020 Framework Programme, under ERC Advanced Grant No. 692670 ``FIRSTORM'’ and  SISSA/CNR project "Superconductivity, Ferroelectricity and Magnetism in bad metals" (Prot. 232/2015). A.V. acknowledges financial support from the Austrian Science Fund (FWF) through the Erwin Schr\"odinger fellowship J3890-N36. C.G. acknowledges support from Universit\`a Cattolica del Sacro Cuore through D1, D.2.2 and D.3.1 grants. M.C. and C.G. acknowledge financial support from MIUR through the PRIN 2015 program (Prot. 2015C5SEJJ001). 
\end{acknowledgments}


\appendix

\section{Non-equilibrium steady-state current}\label{app:NESScurrent}

In our one-dimensional tight-binding model \eqref{eq:tight-binding equation}-\eqref{eq:LopRec}, the non-equilibrium steady-state current (NESS) can be derived from the rate of change of the occupation number $\hat n_N = \ketbra{N}{N}$ on the $N^{th}$ site,
\begin{align}
	\dot{n}_N = \tr{\hat n_N \dot{\hat \rho}}.
\end{align}
Evaluated in the steady state $\hat \rho_\textrm{ss}$, the above rate of occupation change (trivially) vanishes. From the dynamical equation \eqref{eq:tight-binding equation} together with the Lindblad operators \eqref{eq:LopInOut}-\eqref{eq:LopRec} we then obtain
\begin{align}\label{eq:Processes}
	\tr{\hat n_N \dot{\hat \rho}_\textrm{ss}} = &-\frac{i}{\hbar}\tr{\hat n_N [\textrm{H}, \hat\rho_\textrm{ss}]} + \tr{\hat n_N \mathcal{L}_\gin(\hat\rho_\textrm{ss})}\nonumber \\&+ \tr{\hat n_N \mathcal{L}_\gout(\hat\rho_\textrm{ss})} + \tr{\hat n_N \mathcal{L}_\phi(\hat\rho_\textrm{ss})}\nonumber\\&+ \tr{\hat n_N \mathcal{L}_\grec(\hat\rho_\textrm{ss})} = 0. 
\end{align}
One can now associate a different physical process with each one of the  terms in the r.h.s of  Eq.~\eqref{eq:Processes}. The quantity of interest for us is the number of excitons leaving the system through the lead which is captured by the NESS current
\begin{align}
	I := \tr{\hat n_N \mathcal{L}_\gout (\hat\rho_\textrm{ss})} = \gamma_\gout \matelem{N}{\hat{\rho}_\textrm{ss}}{N}.
\end{align}

\section{Single-particle approximation}\label{app:Many-particles}

\begin{figure}
\center
	\includegraphics[width = 0.9\columnwidth]{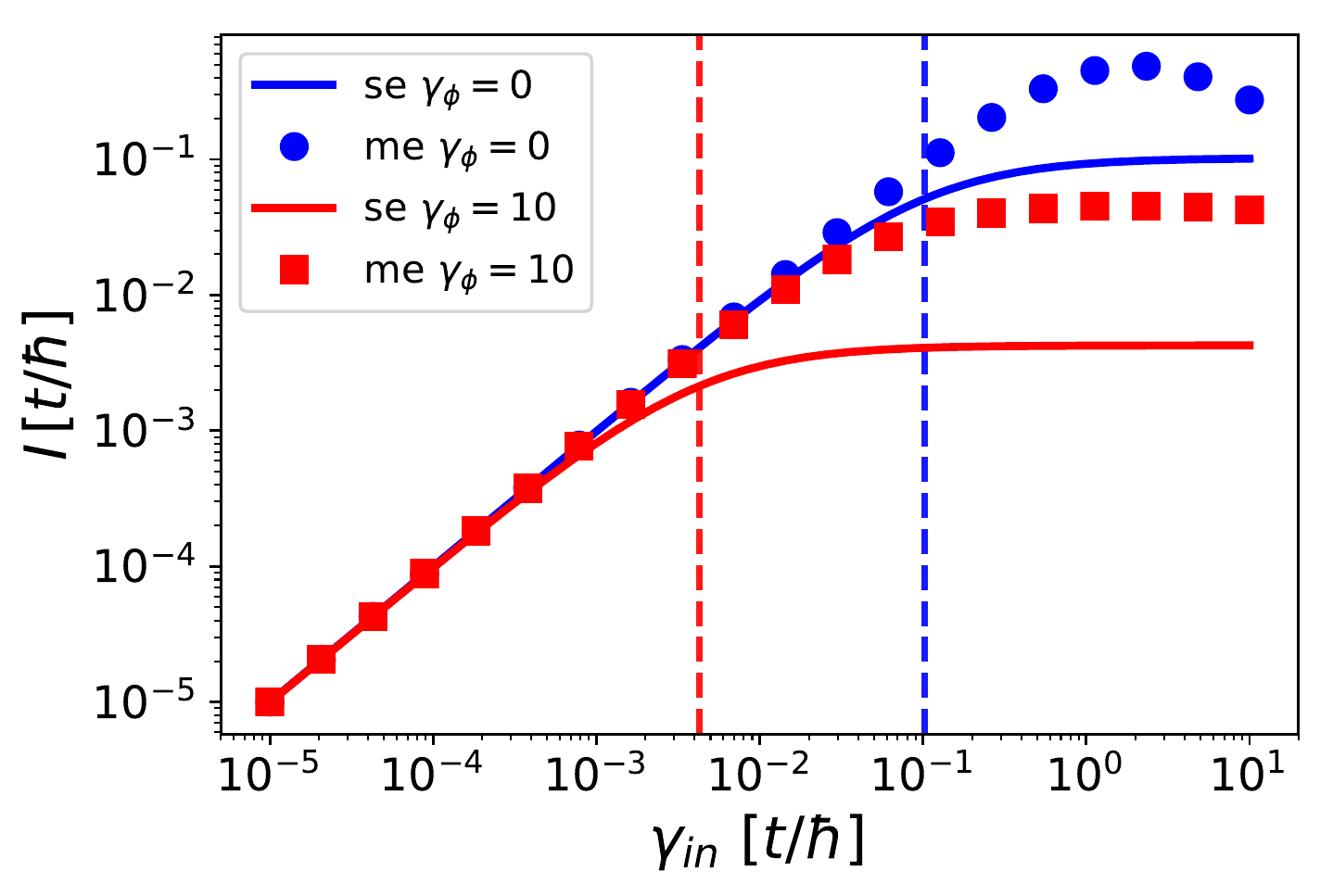}
	\caption{[Color online] Value of the current \eqref{eq:current def} as a function of the pumping $\gamma_\gin$ for zero dephasing (blue curve) and large dephasing $\gamma_\phi$ (red curve) in the single excitation approximation (se) computed from Eq.~\eqref{eq:CurrentAnalytical}.
    Symbols indicate the same current \eqref{eq:current def}, but in the full excitation manifold (me): blue circles for zero dephasing, and red squares for $\gamma_\phi=10$. 
    As one can see, both the single particle approximation and the full excitation calculation give rise to the same current for $\gamma_\gin \leq 1/\tau$, the latter being shown as vertical dashed lines.}
	\label{fig:ManyExcitation}
\end{figure}

The model \eqref{eq:tight-binding equation}-\eqref{eq:LopRec} we use in this manuscript is limited to the single excitation manifold. We argue that this approximation requires the injection rate to be smaller than the inverse of the average transfer time $\gamma_\gin \ll 1/\tau$. To test this limit, we numerically compare the NESS current obtained in our model and the equivalent model within the full excitation manifold. In the latter case, the chain is described by an array of coupled qubits instead of single sites. Then, the Hamiltonian reads
\begin{align*}
	\hat{\textrm{H}} = t \left( \sum_{j = 1} ^ {N-1} \hat\sigma^+_j\hat\sigma^-_{j+1} + \hat\sigma^+_{j+1}\hat\sigma^-_{j}\right),
\end{align*}
and the Lindblad operators are
\ch{
\begin{align*}
	&\hat L_{\phi,j} = \sqrt{\gamma_\phi}\hat\sigma^-_j\hat\sigma^+_j \;\; ; \;\; \hat L_\gout = \sqrt{\gamma_\gout}\hat\sigma_N^- \;\; ; \\& \hat L_\gin = \sqrt{\gamma_\gin}\hat\sigma^+_1 \;\; ; \;\; \hat L_{\grec,j} = \sqrt{\gamma_\grec}\hat\sigma_j^-.
\end{align*}
}
The current is defined as
\ch{
\begin{align}\label{eq:I-many-excitation}
	I_\textrm{me} = \tr{\hat\sigma_N^+ \hat\sigma_N^- \, \mathcal{L}_\gout (\hat\rho_\textrm{ss})} = \gamma_\gout \tr{\hat\sigma_N^+ \hat\sigma_N^- \hat\rho_\textrm{ss}}.
\end{align}
}
\ch{In order to compute the current, we employ the same method as in the single excitation approximation. First, we compute $\rho_\textrm{ss}$ using the steady-state solve of Qutip \cite{Johansson2012}, and then evaluate Eq.~\eqref{eq:I-many-excitation}.} As shown in the example in Fig.~\ref{fig:ManyExcitation}, the single excitation approximation holds for $\gamma_\gin \leq 1/\tau$.

\section{Superradiant transition}\label{app:Superradiant}

The value of the superradiant transition (st) can be derived analytically in the limit of large systems and be computed numerically otherwise \cite{Celardo2009}. In both cases, we consider the linear chain without pumping, dephasing, spontaneous emission and disorder ($\gamma_\gin = 0$, $\gamma_\phi =0$, $\gamma_\grec=0$, $\epsilon_j=0$). It is  convenient to include the decay to the sink as a non-Hermitian term in the Hamiltonian instead of a Lindblad term. In this case, we have to remove the state with no excitation $\ketbra{0}{0}$ from the description and the non-Hermitian Hamiltonian of the system becomes
\begin{align}\label{eq:Non-HermitianHamiltonian}
		&\hat{\textrm{H}}_\textrm{NH} = \hat{\textrm{H}} - \frac{i}{2}\hat{\textrm{Q}} \\& =: t \,\left(\sum_{j \equiv 1} ^ {N-1}  \ketbra{j}{j+1} +\ketbra{j+1}{j} \right) - i\hbar\frac{\gamma_\gout}{2}\ketbra{N}{N},\nonumber
\end{align}
with $\hat{\textrm{H}}$ as defined in Eq.~\eqref{eq:tight-binding equation}. The corresponding Heisenberg equation of motion,
\begin{align}
	\dot{\hat{\rho}}^{\prime}= -\frac{i}{\hbar} \left(\hat{\textrm{H}}_\textrm{NH}\hat{\rho}^{\prime} - \hat{\rho}^{\prime}\hat{\textrm{H}}_\textrm{NH}^\dagger \right),
\end{align}
gives rise to the same dynamics as the master equation \eqref{eq:tight-binding equation} for the density matrix $\hat{\rho}^{\prime}$ which is equal to $\hat\rho$ but without the state $\ketbra{0}{0}$. The dynamics in this case are not trace-preserving. In fact, the asymptotic state has always $\textrm{Tr}(\hat{\rho}^{\prime})=0$ because the excitation leaves the system and ends up in the empty state which we removed from our description.

The superradiant transition happens when one of the eigenstates of $\textrm{H}_\textrm{NH}$ (the superradiant state) takes over all the decay width (i.e., the imaginary part of its eigenvalue becomes larger than that of all others and increases with the coupling to the sink while the others decrease with the coupling, see Fig.~\ref{fig:OptimalCurrent}(b) of the main text). Note that we have only one superradiant state because there is only one open decay channel in our model. In general, there is one superradiant state for each open decay channel.

In order to numerically identify the value of $\gamma_\gout$ at which the transition occurs one compares the imaginary part of the largest eigenvalue of the associated Liouvillian to the average value of the imaginary parts of all other eigenvalues \cite{Celardo2009}. For small rates, all eigenvalues grow with $\gamma_\gout$. At the transition point, the largest value incurs a non-linear transition and increases suddenly, whereas all other eigenvalues have a maximum and begin to decrease. This transition point can thus be identified as the value of $\gamma_\gout$ where the average over all (the imaginary parts of the) eigenvalues, excluding the largest one, has a maximum (c.f. Fig.~\ref{fig:OptimalCurrent}). 
This criterium allows for a very simple and effective numerical way to identify the superradiant transition \cite{Celardo2009}. 

Another, equivalent characterization of the superradiant transition makes use of the fact that at some strength of $\gamma_\gout$ the energy levels will start to overlap with one another. This happens when the average level spacing $D$ and the average decay width  $\langle \Gamma\rangle$ are roughly equal \cite{Celardo2009} (in other words, the point where perturbation theory breaks down). We can then use perturbation theory to approximate this simple intuitive estimate analytically \cite{Celardo2009}. For our system \eqref{eq:Non-HermitianHamiltonian}, the strength at which the transition occurs converges to $\gamma^\textrm{st} \approx 4t/\hbar$ in the limit of long chains ($N\gg 1$) in the absence of losses and disorder. This is consistent with the numerical calculation for which we obtain $\gamma^\textrm{st} = 2t/\hbar$ (c.f. Fig.~\ref{fig:OptimalCurrent}(b) of the main manuscript). 

In the limit of weak decay, i.e., $\gamma_\gout \ll D$, $\hat{Q}$ can be considered as a perturbation with $\gamma_\gout$ the perturbation parameter. The unperturbed energy levels obtained by diagonalization of the system's Hamiltonian $\hat{\textrm{H}}$ and the corresponding eigenvectors read
\begin{align*}
	&\omega_q = 2t\cos\left(\frac{\pi q}{N+1}\right) \\&\ket{q} = \sqrt{\frac{2}{N+1}}\sum_{j=1}^N \sin\left(\frac{jq\pi}{N+1}\right)\ket{j},
\end{align*}
with $q = 1,\ldots,N$. In the limit $N\gg1$, the average level spacing is thus approximately given by
\begin{align*}
	D \approx \frac{\omega_1 - \omega_N}{N} \approx \frac{4t}{N}.
\end{align*}
From standard non-degenerate perturbation theory (c.f. for example the textbook \cite{Sakurai2011}), we have
\begin{align}
	E_q &= \omega_q + \matelem{q}{-\frac{i}{2}\hat{Q}}{q} + \mathcal{O}(\gamma_\gout^2) \nonumber\\& \approx \omega_q -\frac{i\hbar\gamma_\gout}{N+1}\sin^2\left(\frac{Nq\pi}{N+1}\right) =: \omega_q - \frac{i}{2}\Gamma_q \label{eq:DecayRatePerturb}
\end{align}
In other words, due to the coupling to the sink contained in $\hat{Q}$, the energy levels $\omega_q$ acquire a decay width $\Gamma_q$ which is by definition given by the imaginary part of the corresponding eigenvalue of $\hat{Q}$. The later is given by Eq.~\eqref{eq:DecayRatePerturb} when $\gamma_\gout \ll D$. 
In the limit of large $N$, we have $\langle\sin^2\left(\frac{Nq\pi}{N+1}\right)\rangle = 1/2$ and thus the average decay width converges to $\langle\Gamma_q\rangle = \hbar\gamma_\gout /(N)$. 

Finally, the superradiant transition is obtained when the average decay width is equal to the average level spacing, i.e., $\langle\Gamma_q\rangle / D = \hbar\gamma_\gout /(4t)=1$, and reads
\begin{align}\label{eq:SuperradiantTransition}
	\gamma^\st \approx \frac{4t}{\hbar}
\end{align}
We recall that the value in Eq.~\eqref{eq:SuperradiantTransition} differs quantitatively from the numerical solution $\gamma^\st \approx \frac{2t}{\hbar}$. This is because we employ a qualitative criteria, namely $\langle \Gamma \rangle = D$, to identify the superradiant transition, and, in addition, we use perturbation theory to approximation this criteria in a parameter range where perturbation theory is expected to break down. Nevertheless, Eq.~\eqref{eq:SuperradiantTransition} can be used to qualitatively identify the superradiant transition.

\section{Analytical solution for the average transfer time}\label{app:AnalyticTransferEfficiency}

The average transfer time $\tau$ defined in Eq.~\eqref{eq:TauDef} \ch{can be reduced to an analytical form without any time-integral} by direct diagonalization of the Liouvillian $\mathcal{L}$, the superoperator defined as $\dot{\hat{\rho}} = \mathcal{L} \hat{\rho}$.  The dynamics are thus given by $\hat{\rho}(t^{\prime}) = e^{\mathcal{L}t^{\prime}}\hat{\rho}(0)$. Diagonalization of $\mathcal{L}$ yields
\begin{align*}
	\mathcal{L} = V D V^{-1} \;\; ; \;\; D = \sum_{n} E_n |n\}\{n|,
\end{align*}
with $\mathcal{L} |n\} = E_n |n\}$ and thus 
\begin{align}\label{eq:Rho(t)_analytical}
	\hat{\rho}(t^{\prime}) = V \left( \sum_n e^{E_nt^{\prime}} |n\}\{n|\right)V^{-1}\hat{\rho}(0).
\end{align}
Inserting Eq.~\eqref{eq:Rho(t)_analytical} into the definition of $\tau$, \eqref{eq:avg-transfer-time}, and noting that $\mathbb{R}(E_n) < 0 \;\; \forall n$ since we only have one unique decay channel, yields
\begin{align}\label{eq:tau-no-integral}
	\tau = \frac{\gamma_\gout}{\hbar \eta} \{N| V \left(\sum_n\frac{1}{E_n^2} |n\}\{n|\right)V^{-1}|N\}.
\end{align}
Note that when there are no losses in the system, i.e., $\gamma_\grec =0$,  the efficiency is $\eta=1$ because in our model the excitation then invariably leaves the system (as long as $\gamma_\gout > 0 $). \ch{Eq.~\eqref{eq:tau-no-integral} was evaluated numerically to obtain the results for $\tau$ in this manuscript.}

\section{Disorder and losses}\label{app:Disorder}

\begin{figure}
	\centering
	\includegraphics[width = 0.95\columnwidth]{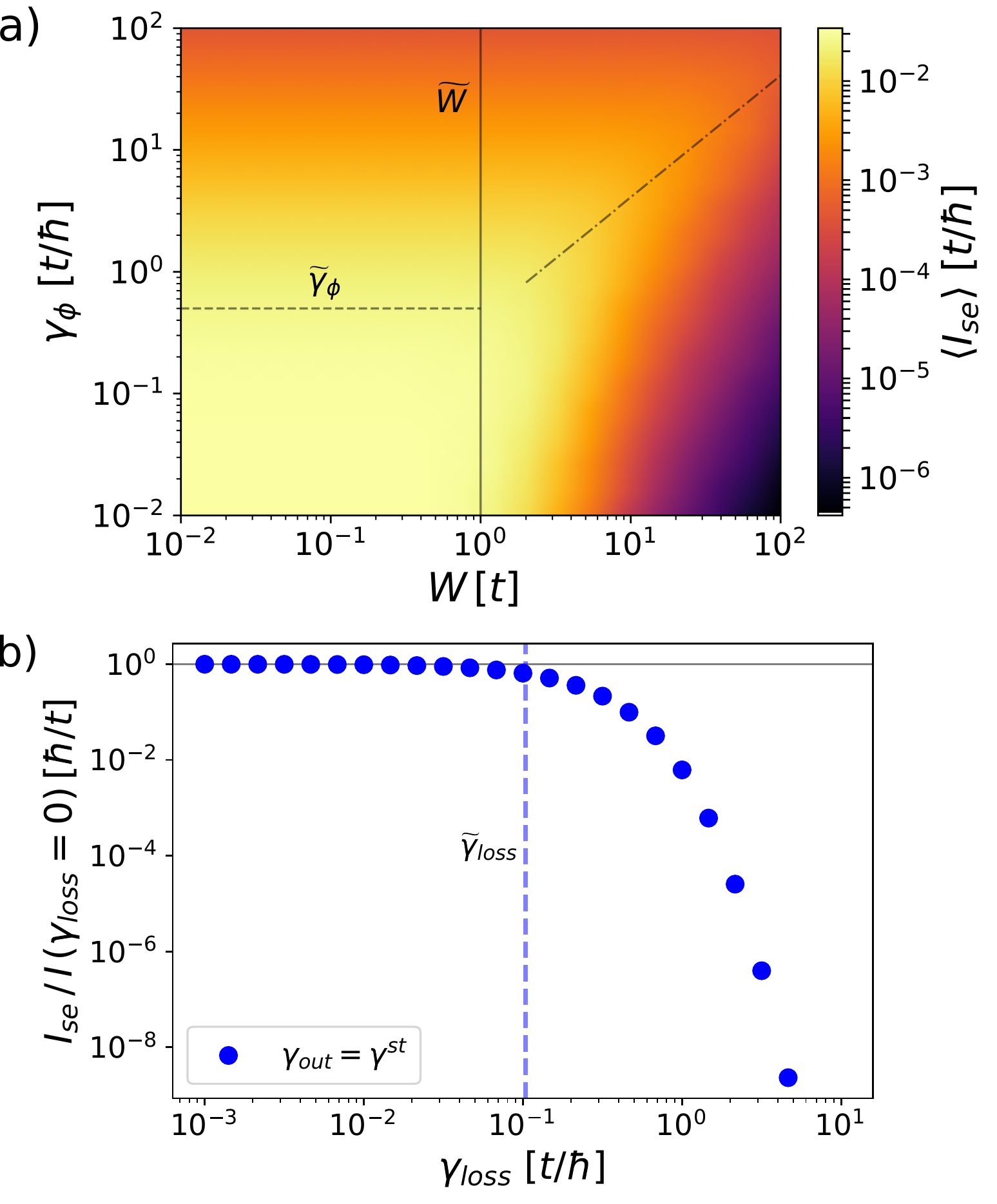}
	\caption{[Color online] Effect of disorder and losses on the maximal current. a) Ensemble averaged maximal NESS current $\langle I_\Imax \rangle$ as a function of the disorder strength $W$ and the dephasing $\gamma_\phi$ at the superradiant transition $\gamma^\st$ obtained from averaging over $100$ realizations of the disorder. For $W < \crit{W} = t$ (left of the vertical full line), the current is independent of $W$, and maximal and independent of the dephasing for $\gamma_\phi \leq \crit{\gamma}_\phi$ (below the horizontal dashed line). For $W>  \crit{W} = t$, we see the dephasing-assisted transport regime: the current is suppressed by localization and is maximal at $\gamma_\phi \approx W / (\sqrt{6}\hbar)$ (diagonal dashed-dotted line). b) Current as a function of the losses at $\gamma_\phi = 0$ and $\gamma_\gout = \gamma^\st$. For $\gamma_\grec \geq \crit{\gamma}_\textrm{loss}=\tau^{-1}|_{\gamma_\grec = 0}$ (vertical dashed line) the current is strongly suppressed.}
	\label{fig:Disorder}
\end{figure}

Let us consider how the introduction of (static) disorder and losses modifies the NESS current $I$. We first focus on the former and consider uniform disorder in the on-site energies $\epsilon_j = \epsilon_j^0 + \delta w$ with $\delta w$ randomly distributed in  $[-W/2,W/2]$. It is known that for one-dimensional chains, in the thermodynamic limit (infinitely long chains), any level of disorder leads to Anderson localization \cite{Anderson1958} which completely inhibits transport. In this case, dephasing can help to overcome localization thus  leading to the so-called dephasing-assisted transport \cite{Plenio2008}. In finite-size chains coupled to a sink, three distinct transport regimes \cite{Zhang2017b} can be observed. For strong disorder ($W \gg t$) localization and dephasing-assisted transport is observed with the optimal dephasing given by $\gamma_\phi = W / (\sqrt{6}\hbar)$ \cite{Zhang2017b}. For medium disorder ($W < t$)  transport is possible even without dephasing whenever the localization length is larger than the chain's length. For chains of medium length ($N\gtrsim100$) an optimal dephasing value still exists when  $t>W>\hbar\gamma_\gout/\sqrt{N}$ \cite{Zhang2017b}. Since in our case we are considering  short chains (in order to keep the transport coherent) and low static disorder ($W\ll t$), transport is mostly unaffected and dephasing always diminishes the current as discussed previously (c.f. Fig.~\ref{fig:OptimalCurrent}(a)).

  Interestingly, the coherence-enhanced current occurring close to the superradiant transition, $\gamma^\st \approx 2 t / \hbar$, is robust against disorder when below the localization threshold. In fact, when the disorder is small compared to the hopping term the current remains completely unaffected as shown in Fig.~\ref{fig:Disorder}(a). Also, in this case, the dephasing is always detrimental to transport and the dephasing $\crit{\gamma}_\phi$ below which the current is optimized at $\gamma^\st$ is given by Eq.~\eqref{eq:MaxDephasing1} (see Fig.~\ref{fig:Disorder}(a)).

Even if in our case average transfer times are much less than realistic recombination times, it is important to discuss its effect. The introduction of losses (i.e., $\gamma_\grec >0$) has the trivial consequence that at a fixed rate in, $\gamma_\gin$, the absolute value of the current is lowered. However, the effect becomes significant only when the losses become comparable or larger than the inverse transfer time in the absence of losses (see Fig.~\ref{fig:Disorder}(b)). 
\begin{align}
	\gamma_\grec \leq \tau^{-1}|_{\gamma_\grec = 0}
\end{align}
In other words, when excitations leave the chain through the sink before they are lost, the overall current is not modified.

\section{Comparison to transport in conventional materials}\label{app:ConventionalMaterials}

Let us briefly comment on the difference, regarding the characteristic time and the physical processes involved in the extraction of the photo-induced carriers, between the conventional scheme which state-of-the-art photo-voltaic devices are based on (see for instance the one described in Ref.~\cite{Wang2015}) and the coherence-driven devices that we suggest in this work.

In conventional devices, the photo-induced charge collection is based on diffusive charge migration at an effective drift velocity. Namely, once the carriers are generated in the material, they thermalize by releasing most of they energy and relaxing into low-energy and, usually, low-mobility states that shorten the diffusion length of photo-carriers. As a consequence, the transfer time (necessary to reach the collectors) can be very long, thus limiting the efficiency of the photo-conversion process. For instance, LVO thick films for photovoltaic applications present a low mobility (0.1 cm$^2$/Vs) due to the defect-induced trapping of low-energy electronic excitations.

\begin{figure}[b]
	\centering
	\includegraphics[width = 0.9\columnwidth]{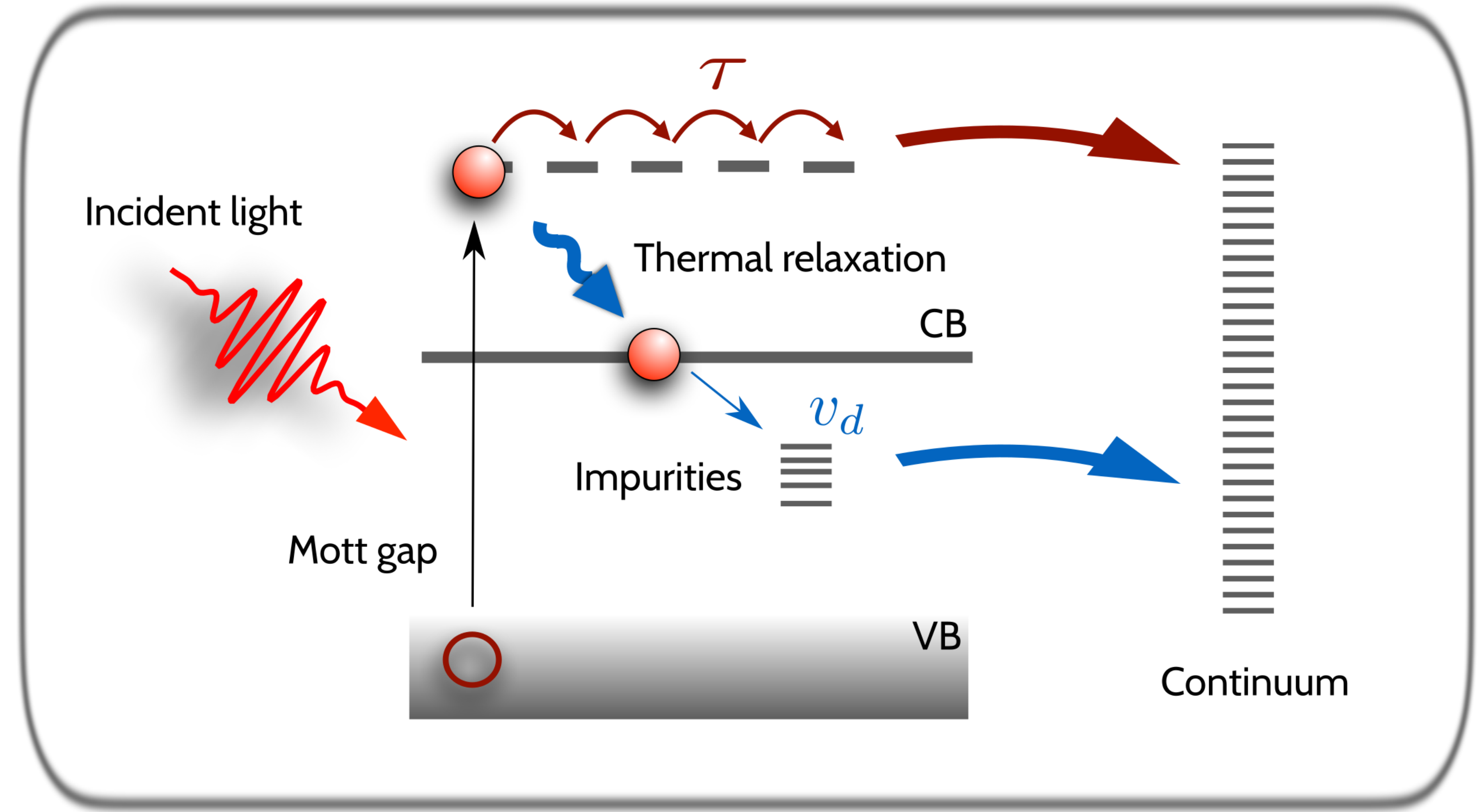}
	\caption{[Color online] In conventional photovoltaic devices, the high-energy excitation created by photo-absorption first thermalizes and then migrates to the electrodes (continuum) with an effective drift velocity $v_d$ and can be trapped in low-energy states created by defects in the structure (blue arrows). In the few-atomic layers devices we consider here the excitation migrates to the electrodes before thermalization occurs in a time $\tau$ that is minimal at the superradiant transition (red arrows).}
	\label{fig:ThermalTransport}
\end{figure}

On the contrary, in the case of the building block we propose, which is made up of few monolayers of material (~5-nm-thick), the photo-generated charge excitations are coherently collected by the electrodes without any intermediate low-energy and low-mobility states along the collection pathway. As a consequence, the efficiency of the process is controlled by the quantum laws. When the transfer time is much shorter than the decoherence time, the current can be optimized thanks to the superradiant phenomenon. The conventional incoherent scheme and the quantum-enhanced path are schematically represented in Fig.~\ref{fig:ThermalTransport}.

\vfill
 

\bibliographystyle{apsrev4-1}
\bibliography{OneDChainCurrent.bib}

\end{document}